\documentclass[a4paper,11pt]{article}

\usepackage{preamble}
\usepackage{xcolor}
\makeindex 
\title{\vspace{-2cm}\Huge{The Hyper-Kamiokande experiment: 
input to the update of the European Strategy for Particle Physics}}

\DeclareUnicodeCharacter{2212}{-}

\hypersetup{urlcolor=black}

\author[1]{K.~Abe}
\author[2]{P.~Adrich}
\author[3]{R.~Ahl~Laamara}
\author[4]{H.~Aihara}
\author[5]{A.~Ajmi\,\orcidlink{0000-0002-0447-7362}}
\author[6,a]{R.~Akutsu}
\author[7]{H.~Alarakia-Charles}
\author[]{I.~Alekseev\,\orcidlink{0000-0003-3358-9635}}
\author[8]{Y.~Alj~Hakim\,\orcidlink{0000-0002-9065-1303}}
\author[9]{S. Alonso Monsalve\,\orcidlink{0000-0002-9678-7121}}
% ...
\affil[1]{Kamioka Observatory, Institute for Cosmic Ray Research, University of Tokyo, Kamioka, Gifu, 506-1205 Japan}
\affil[2]{National Centre for Nuclear Research, ul. Soltana 7, Otwock, 05-400, Poland}
\affil[3]{Faculty of Sciences, Mohammed V University, Rabat}
\affil[4]{University of Tokyo, Department of Physics, 7-3-1 Hongo, Bunkyo-ku, Tokyo, 113-0033, Japan}
\affil[5]{University of Winnipeg, 515 Portage Ave, Winnipeg, Manitoba, R3B 2E9, Canada}
\affil[6]{High Energy Accelerator Research Organization (KEK), 1-1 Oho, Tsukuba, Ibaraki, 305-0801, Japan}
\affil[7]{Physics Department, Lancaster University, Lancaster, LA1 4YB, United Kingdom}
\affil[8]{University of Sheffield, Department of Mathematical and Physical Sciences, Western Bank, Sheffield, S10 2TN, United Kingdom}
\affil[9]{ETH Zurich, Institute for Particle Physics and Astrophysics, Otto-Stern-Weg 5, Zurich, CH-8093, Switzerland}
% ...
\affil[a]{also at J-PARC, Ibaraki, Japan}
% ...
\usepackage{setspace}
\makeatletter
\let\oldaffillist\AB@affillist
\renewcommand{\AB@affillist}{\begin{flushleft}\begin{spacing}{1.5}\oldaffillist\end{spacing}\end{flushleft}}
\makeatother

\begin{document}
%\author{\Large{The Hyper-Kamiokande collaboration\footnote{ Contact information for %corresponding author: Francesca Di Lodovico (King's College London), Hyper-Kamiokande %international co-spokesperson, 
% \href{mailto:francesca.di_lodovico@kcl.ac.uk}{francesca.di$_$lodovico@kcl.ac.uk}.}}
%\begin{picture}(0,0) \put(0,0){\includegraphics[width=2cm]{logo}} \end{picture}
%}

%\input{ESPPU_authors}

%\date{($\rm{16^{th}}$ January 2023)}
\date{}
%\titlepic{\includegraphics[width=0.3\textwidth]{KCLlogo.png}

\maketitle
%\vspace{2ex}
%\bigskip
%\input{authorHK}
%\bigskip
%\vspace{35ex}

\abstract{
This document summarises the input of the Hyper-Kamiokande collaboration to the
2026 Update to the European Strategy for Particle Physics, ESPPU.

Hyper-Kamiokande is  a large infrastructure for particle and astroparticle physics being built in Japan and aiming to start operations by the end of 2027 whose objective is to address the most important questions in science today, for instance how the universe began and evolved. It aims to measure with the highest
precision the leptonic Charge-Parity violation parameter that could explain the baryon asymmetry in the Universe and study / challenge the standard three-flavour neutrino framework using both a Mega-Watt intense neutrino beam and high-statistics atmospheric neutrino samples. The combination of these samples will break the degeneracies between the effects of the Mass Ordering and Charge-Parity violation, allowing for their measurement without relying on external information.

\noindent Hyper-Kamiokande is also a neutrino observatory for astrophysical events that will collect the highest statistics due to its size. It will also be able to precisely measure solar neutrino oscillations and other astrophysics events as supernova bursts, relic supernova neutrinos, neutrinos in correspondence with gravitational waves, etc. It can also detect neutrinos from sources such as dark-matter annihilation, gamma-ray burst jets, and pulsar winds, further increasing our understanding of some of the most spectacular, and least understood, phenomena in the Universe.

\noindent Furthermore, due to its size and particle identification capability, the experiment has an excellent potential to search for proton decay, providing a
significant improvement in discovery sensitivity over current searches for the proton lifetime and nucleon decays.

Hyper-Kamiokande is expected to run at least 20 years from the start of operations and is supported by 10 countries in Europe that are contributing to its construction, future operation and data analysis. Prototyping and assembly are also being carried out at CERN. 
The reduction of the flux systematic uncertainties would benefit from new hadron production measurements at the NA61/SHINE experiment at CERN, also with a low-energy beam. 
A final upgrade of the magnetised off-axis near detector (ND280++) for the high-statistics phase in the 2030s aim to be sought and would benefit from CERN support.

%\thispagestyle{empty}
%\pagenumbering{roman}
%\setcounter{page}{1}
\newpage
%\tableofcontents
%\newpage
\pagenumbering{arabic}
\setcounter{page}{1}
\captionsetup{width=\linewidth}
\section{Executive Summary}
\vspace{0.3cm}
\label{sec:executivesummary}
Hyper-Kamiokande (\hkshort), based on the experience from the Super-Kamiokande (\skshort)~\cite{SK} and T2K\cite{T2KExperiment} experiments, is being constructed by the international physics community to provide a wide and groundbreaking multi-decade physics programme from 2027.

Within the next 3 years, \hkshort will complete the construction of the far and near detectors, the upgrade of the neutrino beam and will start operations.  Contributions from European institutes have been crucial for all the aspects of the infrastructure. 

The science that will be developed will focus on measuring the Charge-Parity (CP) violation in the leptonic sector, to precisely test the three-flavour neutrino oscillation paradigm and search for new phenomena,
using a Mega-Watt (MW) accelerator beam. The large sample of atmospheric neutrino will provide a sensitivity to the determination of the neutrino Mass Ordering (MO) between 4 and 6 standard deviations.
A wide astrophysical programme will be carried out at the experiment that will also allow us to measure precisely the solar neutrino oscillations. A set of other physics searches is planned, such as indirect dark matter. 
\hkshort also has an excellent ability to search for proton decay, providing a significant improvement in the discovery sensitivity over current searches for the proton lifetime of two orders of magnitude. 
If CP is maximally violated, \hkshort could discover it within $3$-$4$ years from the beginning of data taking, if MO is known or for non-degenerate values of \dcp -MO.
The CP violation discovery potential for small values of \dcp and the precise measurement of the neutrino oscillation parameters will be more affected by systematic uncertainties, mainly from a potential mismodeling of the electron neutrino-to-antineutrino event rate.

Thus, a programme dedicated to further reduction of the neutrino flux systematic uncertainties would profit greatly from the support of auxiliary hadron production measurements, such as those at the NA61/ SHINE experiment at the CERN Super Proton Synchrotron (SPS)~\cite{NA61:2014lfx} including also the realisation of a new beamline~\cite{Nagai:2810696} at CERN for low-energy beam.
In view of the high-statistics phase, after 2030, when neutrino cross-sectional systematic uncertainties will have become dominant, \hkshort is preparing for a possible final upgrade of ND280, called ND280++, for which the Neutrino Platform at CERN will be indispensable.

\section{Introduction}
\vspace{0.3cm}
\label{sec:introduction}

%{\em{Francesca}}

\noindent

% \tbd{
% \begin{itemize}
%     \item Overview 
%     \item Connect to previous EU strategy.
%     \item Describe the activity of the group and highlight EU contributions.\\
% \end{itemize}
% }

%
% Brief physics context and HK physics program
%

% On the strength of a double Nobel prize winning experiment (Super)Kamiokande and an ex- tremely successful long baseline neutrino programme, the third generation Water Cherenkov detec- tor, Hyper-Kamiokande, is being developed by an international collaboration as a leading worldwide experiment based in Japan.

\hkshort is a large infrastructure for particle and astroparticle physics that leverages the successful experience of water Cherenkov detectors (Kamiokande and \skshort, in Kamioka, Japan) with a detector $8.4$ times larger than \skshort, acting both as a far detector of a long neutrino oscillation experiment (LBL) with a \SI{295}{\km} baseline and a neutrino observatory. It also leverages the upgraded $1.3$\,MW neutrino beam from the T2K~ long-baseline neutrino experiment produced at the Japan Proton Accelerator Research Complex (J-PARC), and its suite of near detectors (ND280) at \SI{280}{\meter} from the neutrino source, which will be complemented by a new Intermediate Water Cherenkov Detector (IWCD) at \SI{850}{\meter} from the source. The infrastructure is hosted by Japan (KEK and University of Tokyo).
%\hkshort has the same baseline as that of T2K and a far detector 8.4 times larger than \skshort (258 kilotonne water-Cherenkov).  \hkshort 's main goal is the precise measurement of the neutrino flavour oscillation probability to search for the violation of the Charge-Parity (CP) symmetry in the leptonic sector, violation that could help to unravel the matter-antimatter asymmetry in the universe. The far detector will be also act as an unparalleled observatory in which a wide range of physical properties of the Universe will be investigated. They include searches for proton decay, a process not yet observed; the most accurate measurement of neutrinos from supernovae; and the first evidence of relic neutrinos from supernovae in the early Universe. Unique physics and discovery potential will be provided from 2027. 
%\textbf{Physics programme:}
%Hyper-Kamiokande (\hkshort) is the next-generation 
%neutrino-oscillation
%\tbd{water-Cherenkov} 
%\DS{or ``water-Cherenkov'' ?}
%experiment in Japan for 

Hyper-K's main goal is the precise measurement of the neutrino flavour oscillation probability to search for the violation of CP symmetry in the leptonic sector, which could help to address one of the main questions in science that is the matter-dominated Universe.
%from an initial balance of particle and antiparticles at the Big Bang. 
%\DS{We should be careful with this last sentence to not oversell low-energy CPV for leptogenesis as well as with the initial unbalance}
\hkshort can measure the CP complex phase (\dcp) with a resolution between $6^{\circ}$ and $20^{\circ}$, depending on its true value, and determine the neutrino MO combining accelerator and atmospheric neutrinos.  
%One of its main goals is the precise measurement of neutrino oscillations to search for Charge-Parity (CP) violation in the leptonic sector. 
\dcp is directly determined by precisely measuring the \(\nu_{\mu} \rightarrow \nu_e\) and \(\bar{\nu}_{\mu} \rightarrow \bar{\nu}_e\) oscillation channels and comparing the corresponding neutrino interaction rates as a function of the energy at the near- and far-detectors.
A value of \dcp different from 0 and $\pi$ changes the \(\nu_{\mu} \rightarrow \nu_e\) and \(\bar{\nu}_{\mu} \rightarrow \bar{\nu}_e\) oscillation asymmetrically.
The degeneracy between \dcp and MO will be resolved by the unprecedentedly high-statistics sample of atmospheric neutrinos that \hkshort will be able to collect allowing it to achieve the world-best sensitivity to the discovery of CP violation without relying on any constraint on the MO from external experiments.

As neutrino observatory, \hkshort will provide sensitivity to several important phenomena. It will be able to precisely study solar neutrinos including solar neutrino upturn and day-night effects. It will be able to measure supernova burst neutrinos and provide important information on their characteristics including rate, flavour, time, and energy resolution to be able to distinguish between different models of supernova explosion. 
It will be able to search for diffuse supernova neutrino background and study other astrophysical phenomena in the Universe, and also search for indirect evidence of dark matter.
%through the observation of annihilation\CG{fix this sentence. Is this WIMP annihilation in the Sun?}
%Indirect detection searches for the products of WIMP annihilation or decay. 
%\DS{Is this sentence broken ?}
%This is generally done through observations of $\gamma$-ray photons or cosmic rays, gamma ray burst jets, and pulsar winds, etc.
%indirect 

%The \hkshort physics program is even broader.
\noindent With this huge detector mass, \hkshort will search for proton decay, the smoking gun of Grand Unified Theories (GUTs)~\cite{Georgi:1974sy}, surpassing existing \skshort limits after a few years of data taking. 
%Among other channels, \hkshort will have the best sensitivity to proton decay in the $p \rightarrow e^+ \pi^0$ channel, reaching $6\times 10^{34}~yrs$ after 10 years of data taking, and in the $p \rightarrow \nu K^+$, reaching $2\times 10^{34}~yrs$ after 10 years.

\noindent Physics beyond the standard model (BSM) also can be tested, both at the near and far detectors or their combination, for instance feebly interacting particles (FIP), heavy neutral leptons, sterile neutrinos, the neutrino magnetic moment, neutrino non-standard interactions, axions, etc.
\hkshort  provides a state-of-the-art, high-precision infrastructure that will be operated by scientists from $22$ countries to address fundamental scientific questions during at least $20$ years from 2027.  It is supported by $\sim 350$ European members, corresponding to more than $50$\% of the collaboration. European institutes were among the pioneering institutes, that led to the formation of the \hkshort collaboration in 2020 (and continuing from the \hkshort proto-collaboration founded in 2015). This continues and leverages the work by European institutes that started more than two decades ago working in K2K~\cite{K2K} and, later, T2K and \skshort.
Within \hkshort,  European institutes have been contributing to the design, R\&D, and currently  construction of the far detector; the operation and recent upgrade of ND280, within the T2K experiment, that will continue its operations during the \hkshort era; and the design and ongoing construction of IWCD.

The importance of \hkshort was highlighted in the 2020 Update of the European Strategy for Particle Physics (ESPPU)~\cite{CERN-ESU-015}: \textit{Europe, and CERN through the Neutrino
Platform, should continue to support long baseline experiments in Japan and the
United States}.

\section{Hyper-Kamiokande experiment}
\label{sec:experiment}
In the following, we will describe the three facilities of the infrastructure highlighting the European contributions.
\subsection{The water-Cherenkov far detector}
\label{sec:far-detector}
The construction of the new massive water Cherenkov far detector, \hkshort, that serves as a far detector in the long-baseline measurements, and neutrino observatory, located in the largest human-made cavern, in the Tochibora mine at the Kamioka facility, is underway.
Photographs of the cavern are shown in Fig.~\ref{fig:cavern-excavation}.
%in Japan.
The far detector, shown in Fig.~\ref{fig:far-detector}, consists of a cylindrical-shape tank detector, \SI{69}{\meter} in diameter and \SI{71}{\meter} in height, filled with $258,000$ metric tons of ultrapure water. With a fiducial mass of approximately $184$\,kilotonnes,  eight times larger than that of \skshort, it will be the largest underground water tank ever built.
\begin{figure}[!htb]
\centering
    \includegraphics[width=8.cm]{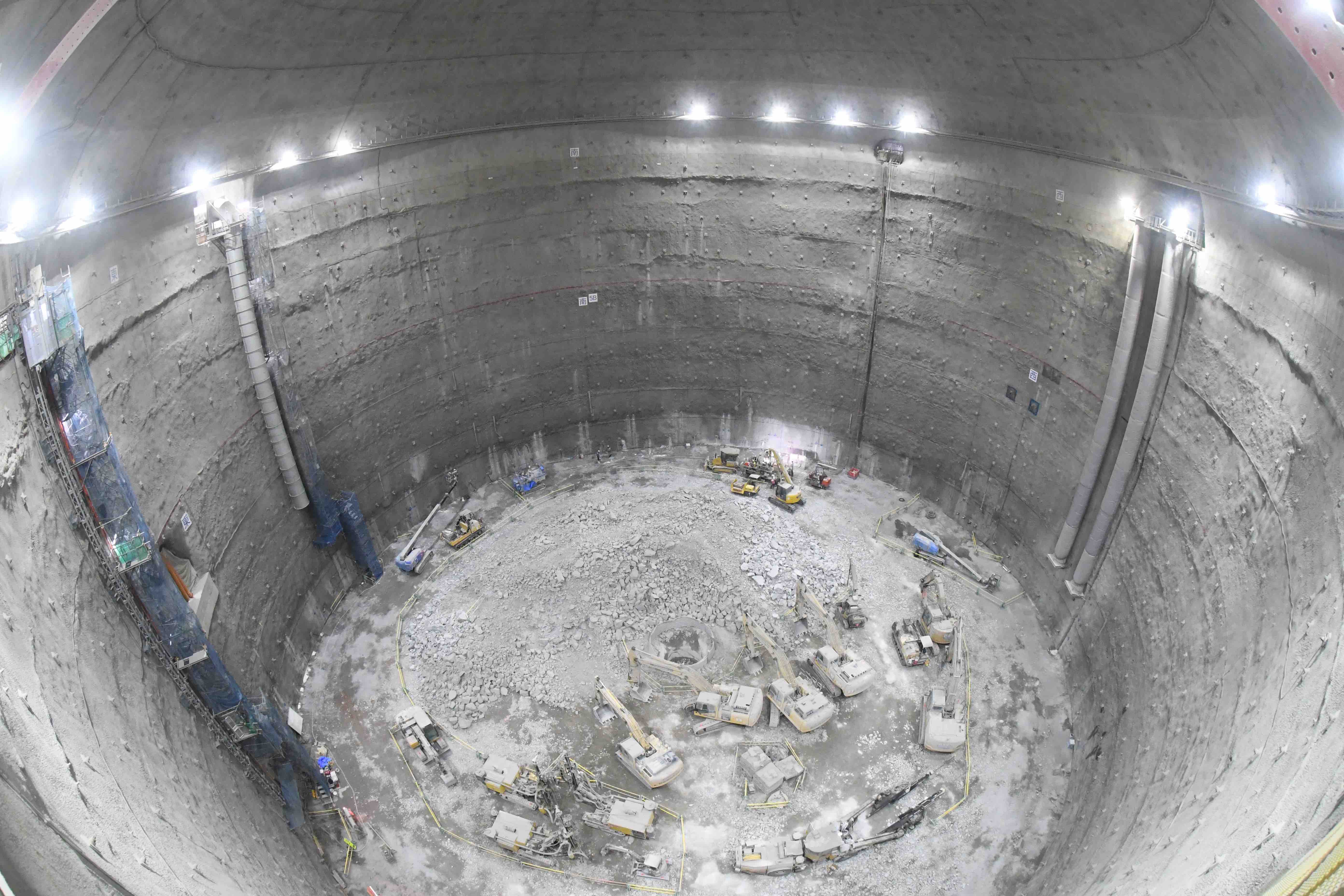}
    \includegraphics[width=8.cm]{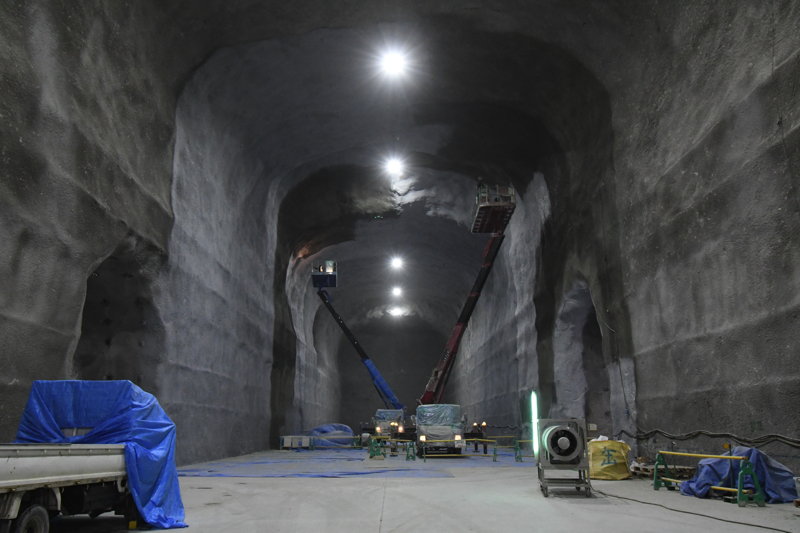}
    \caption{
    Photographs of the \hkshort cavern during the excavation in September 2024. Left: the main cavern being excavated to a depth of \SI{45}{\meter} in the barrel section. Right: the water purification system room under dust-proof coating.
    }
    \label{fig:cavern-excavation}
\end{figure}

The Cherenkov light emitted by charged particles will be read out by photomultiplier tubes (\pmt). 
The detector is divided into two optically separated parts. 
The ``Inner Detector'' (ID) will provide the target mass for the GeV and MeV neutrino interactions, as well as the volume sensitive to the products of proton decay.
The performance of the ID photodetection system exceeds that of \skshort. 
%It adopts two types of photosensors, 
It consists of approximately \NpmtIDApprox $20$-inch diameter PMTs and \Nmultipmt~multi-PMTs modules pointing inward to cover $20$\% of the ID surface.
The $20$-inch PMTs are the same size as in \skshort, with twice the photodetection efficiency ($40$\%)
and time resolution (\SI{1.5}{\ns})
while the dark count rate ($4$ kHz) is similar \cite{Bronner:2020ibs}.
The improved performance will enhance the reconstruction of the particle momenta %resolution \tbd{by a factor two} 
and the spatial resolution of the neutrino interaction vertex. The $20$-inch PMTs are put inside covers to avoid the very rare possibility that a chain implosion may occur during the $20$ or more years underwater. %initiated by the explosion of one PMT.
The ID system will also comprise
\Nmultipmt~multi-PMTs, composite photosensors made of $19$ $3$-inch PMTs that also embedded the front-end electronics. 
Owing to a more precise reconstruction of the particle direction,
%as well as an even lower time resolution (1.3~ns)
%The complementarity of 
the multi-PMTs complement the 20-inch PMTs and allow the two systems to cross calibrate each other. %reducing the ovdetector systematic uncertainties.
The ``Outer Detector'' (OD) consists of about 
%\NpmtOD~
\NpmtODApprox outward-looking 3-inch diameter PMTs coupled with wavelength-shifting plates to increase the overall light yield. The OD overall light collection is also maximised by high-reflectivity white Tyvek on the cavern wall and the outside of the tank. The OD acts as a veto from cosmic rays and generally external background and helps to identify neutrino interactions partially contained in the ID.

The contributions of European institutes are crucial for the realisation of the far detector and include the development, production, and assembly of multiple components.

\noindent It includes PMT cover production, which have been tested at high pressure in the Mediterranean Sea. They are produced in Europe and then shipped to Japan for installation.

\noindent Furthermore, $75$\% of the multi-PMTs are produced and tested in Europe, including at CERN as part of the Water-Cherenkov Test Experiment (WCTE); see Section~\ref{sec:near-detectors} for more details.
The OD photosensing system has been designed in Europe, where most of the tests have also been performed. 
%\DS{Add a couple of sentences with details about where things are done?}
%
\begin{figure}[!htb]
\centering
    \includegraphics[width=14.cm]{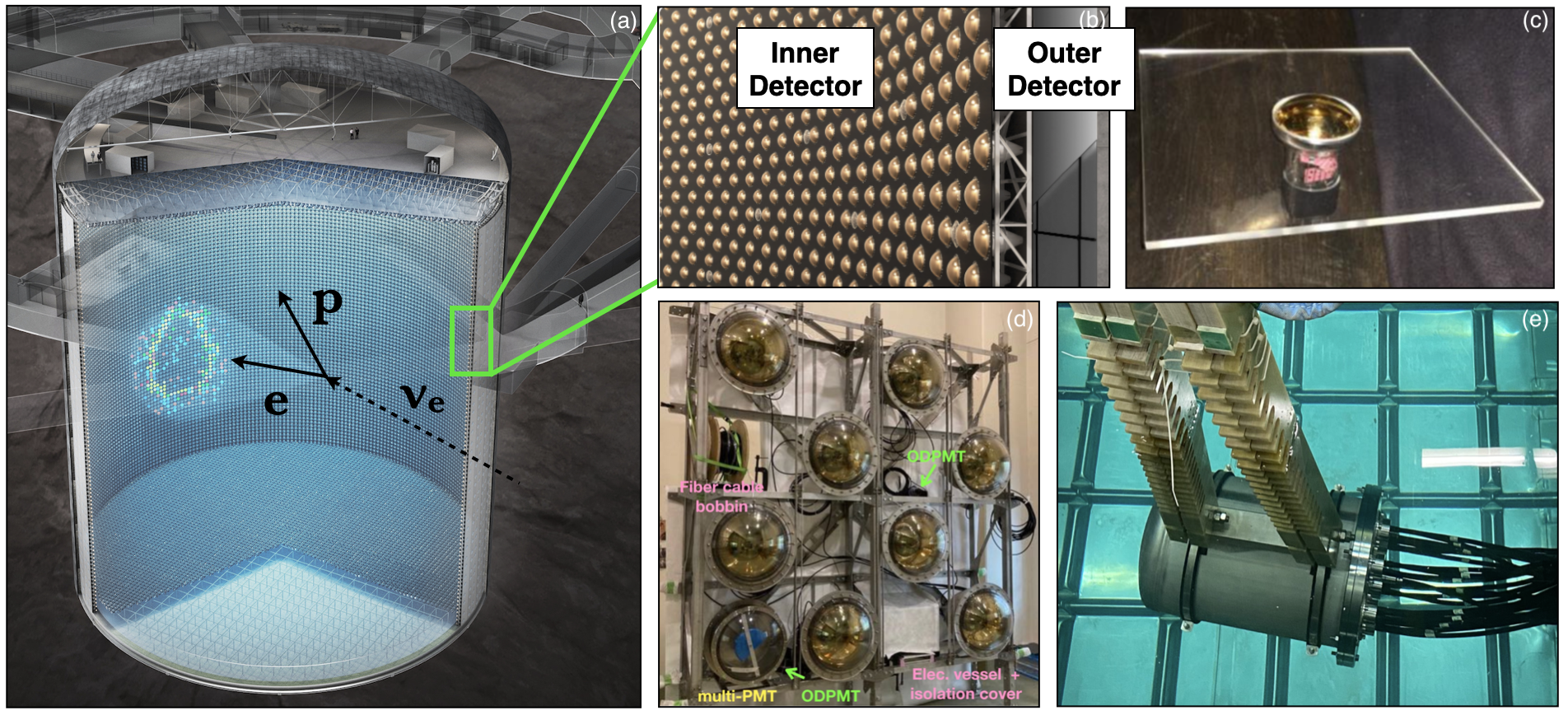}
    \caption{
    (a): Model of the water-Cherenkov far detector.
    (b): Model of the ID and OD \pmt system. 
    (c): Photograph of a OD $3$-inch \pmt{} coupled with a wavelength shifting plate. 
    (d): Mockup of the installation frame with ID and OD \pmt{s} installed (Kashiwa, Japan).
    (e): Electronics unit installed in water at the WA105 cryostat tank at CERN.
    %\DS{replace with higher-quality photo}
    }
    \label{fig:far-detector}
\end{figure}
\noindent European institutes also lead the efforts for the realisation
%development and the production 
of the far-detector readout electronics system for the ID and OD systems. 
\SI{20}{\meter} long signal and high voltage cables digitise the analogue signal from the PMTs and transmit it to the Data Acquisition System (DAQ) system on top of the tank through \SI{150}{\meter} long optical fibres. As the electronics are underwater, the electronics boards are contained inside a waterproof canister able to stand a pressure up to $8$\,bar. 
%for the 20-inch ID \pmt{s} and the 3-inch OD \pmt{s}.
There are a total of about \nunits underwater electronics units (\underwaterunitshort{s}) containing ID and OD electronics.
%(including spares) 
They are synchronised at the \SI{100}{\ps} level by a time generation and distribution system that sends the clock to each digitiser board.  
The assembly and quality assurance, including high-pressure in-water tests, of each \underwaterunitshort will be performed at CERN throughout 2026 (NP08~\cite{Botao:2867639,Botao:2887744})
%in EHN1 over 2026 with the support of the Neutrino Platform, within the NP08 project approved in September 2024 \cite{CERN-DG-RB-2024-531}, 
before being shipped to Japan to be installed in the far-detector tank \cite{Botao:2867639,Botao:2887744}.
%\tbd{FDL: we need to briefly also add the contributions that didnt need CERN/}
European contributions also include the geomagnetic field compensation system to mitigate the  effects of Earth’s magnetic field  on the $20$-inch PMTs. 
\textbf{Ensuring the continuing support of Europe to the construction of the far detector is crucial to the realisation of the \hkshort experiment.}

\noindent The \hkshort DAQ software has been developed in Europe, and will continue to be managed throughout the life of the experiment.

\noindent Finally, the calibration system is a multifaceted system as it needs to deal with different types of calibration to satisfy several needs, over a large energy range from MeV to TeV. It is then composed of a variety of systems, many from Europe, including an accelerator system (LINAC) for low-energy calibration.
\subsection{The neutrino beam}
\label{sec:neutrino-beam}

The neutrino beam is produced at J-PARC by accelerating protons up to \SI{30}{\gev} and directing them onto a \SI{90}{\cm} long graphite target,
producing mainly pions and kaons.
After the upgrade of the J-PARC proton accelerator and the neutrino beamline \cite{BeamUpgrade},
T2K is now steadily running with a beam intensity of $0.8$~MW~\cite{giganti_2024_12704703} 
that will increase to $1.2$\,MW at the start of \hkshort and  $1.3$\,MW in 2028 by reducing the loss of the beam with improved optics and the cycle rate up to $0.86$\,Hz. 
%1.16 second cycle
%by upgrading the radio-frequency system.
The higher beam focussing current ($320$\,kA),
%of the beamline, 
achieved with the replacement of the horn power supplies,
increases the neutrino (antineutrino) flux by an additional $10$-$12$\% and reduces the wrong-sign antineutrino (neutrino) contamination.

European institutes are responsible for the design and construction of the neutrino production target, for developing the required upgrades for operation at the increased beam power and for key components of the target station. These items will be delivered and implemented before the start of \hkshort.
In order to reduce systematic uncertainties we will construct a physics
replica target and measure hadron-production cross-section with NA61/SHINE at CERN. More information will be provided in Section~\ref{sec:ultimate-syst-unc}.
Furthermore, there will be other European contributions, such as the 
%Global Positioning system (GPS) 
Global Navigation Satellite System (GNSS)
upgrade, Optical Transition Radiation Monitor (OTR) readout, beamline orbit/optics modelling, which should help with future beam control, as well as flux systematics. 

\subsection{The near detectors: ND280 and IWCD}
\label{sec:near-detectors}

Constraining the neutrino flux and systematic cross section uncertainties down to few \% level is necessary for the discovery of CP violation.
In particular, an accurate modelling of the \nue to \nueb cross section ratio, \sigmanuenueb, is needed as it could bias the measurement of \dcp, and it is currently based on theoretical estimations. We aim to improve 
it to smaller than $4.9$\%, 
the uncertainty currently affecting the search for CP violation at T2K \cite{T2K:2019bcf}. 
The \hkshort long-baseline physics programme will be enabled by a complex of complementary Near Detectors: IWCD and ND280, that will 
measure the charge of the final-state lepton, 
collect high-statistic samples of \nue and \nueb events 
(exploiting the $\sim 1\%$ contamination in the \num / \numb dominated beam), 
measure the neutrino cross section in water (same medium as the far detector), and
precisely reconstruct the hadronic part of the interaction final state to accurately model the extrapolation from \num to \nue cross section.

\textbf{ND280:} comprises an on-axis detector, INGRID, and a off-axis magnetised detector, ND280-OA.
INGRID, made of layers of plastic scintillator bars alternating with iron sheets,
%monitors the stability of the neutrino beam  
monitors the neutrino event rate and the beam width by spanning the $0^{\circ}$-$1^{\circ}$ off-axis range.
\begin{figure}[!htb]
\centering
    \includegraphics[width=3.5cm]{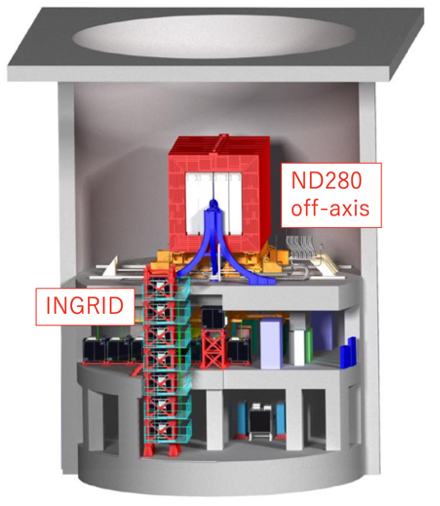}
    \includegraphics[width=7.cm]{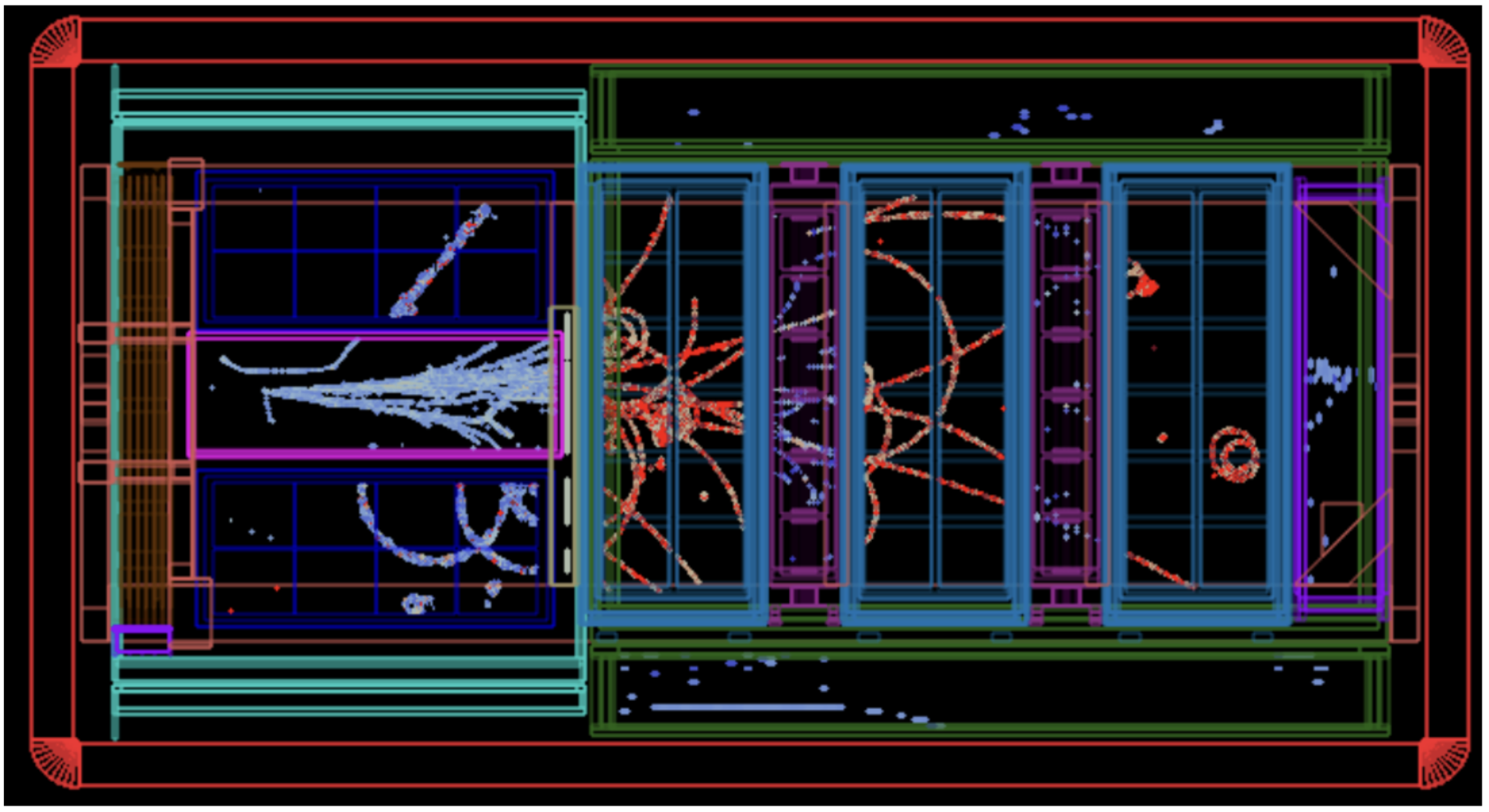}
    \includegraphics[width=6.cm]{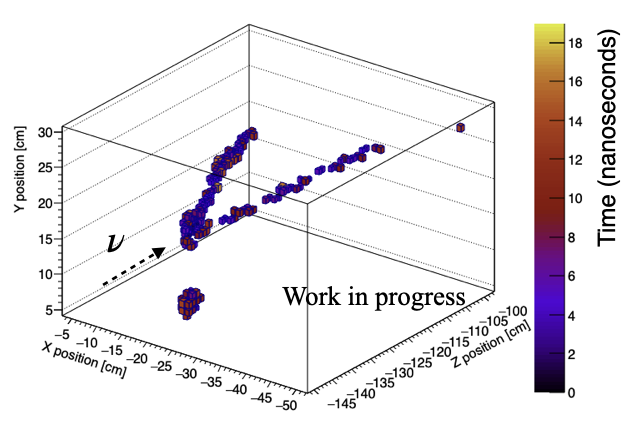}
    \caption{
    Left: 3D model of the ND280 pit. 
    Middle: event  
    %of a \tbd{neutrino charged current interaction} 
    recorded at ND280 during the T2K data taking in June 2024. The upgraded part is within the left box (cyan) with SuperFGD (pink), the two horizontal TPCs (red), surrounded by the TOF.
    Right: a neutrino interaction with a final-state neutron candidate detected in SuperFGD. A neutron kinetic energy of \SI{27}{\mev} is estimated with the time-of-flight technique. The ND280 events are taken from \cite{sfgd-seminar-cern}. 
    %\DS{make sure we can show ND280 event displays. Maybe we can communicate it to the T2K EC}
    }
    \label{fig:nd280}
\end{figure}

ND280-OA is a detector suite situated inside the UA1 magnet, donated by CERN. Two plastic scintillator-based detectors (FGD1 and FGD2) provide $2$~tonnes of active neutrino target with $0.5$ tonnes of water. 
%FGD2 contains about 0.5~tonnes of water, the same medium as at the far detector. 
Three vertical time projection chambers (TPCs) reconstruct the charge and the momentum of forward going particles.
All are surrounded by the electromagnetic calorimeter.
A recent upgrade of ND280, supported by the CERN Neutrino Platform (NP07) \cite{ND280upgrade-tdr,Giganti:2713578},
enables a more accurate reconstruction of neutrino interaction final states including, for the first time, the neutron kinematics.
A $3$D high-granularity plastic scintillator target, called SuperFGD, made of $2,000,000$ optically-isolated 1~cm$^3$ plastic scintillator cubes \cite{Blondel:2017orl}, 
is sandwiched between two horizontal TPCs \cite{Attie:2022smn,tpc-seminar-cern},
%that adopt the resistive micromegas technology
surrounded by a time-of-flight detector (TOF) \cite{Korzenev:2021mny}.
ND280 is currently operating in the T2K experiment (see Fig.~\ref{fig:nd280}) and 
%ND280 is currently collecting neutrino data as near detector of the T2K experiment.
will be operated by \hkshort when it starts to take data.
Before physics operations, both the INGRID readout electronics and the ND280-OA magnet power converter will be refurbished.

%\textbf{Intermediate Water Cherenkov Detector (IWCD):}
\textbf{IWCD:}
With a cylindrical inner detector of \SI{7}{\meter} in diameter and \SI{8}{\meter} in height, IWCD will have total water active mass of $600$\,tonnes and a fiducial mass of about $150$\,tonnes.
%to detect approximately 100,000 \nue and \tbd{30,000 \nueb interactions} in water.
The Cherenkov light is detected by $370$ multi-PMTs, each one 
%with good spatial and timing resolution 
containing $19$ $3$-inch diameter PMTs and the integrated readout electronics.
Moreover, the detector will be moved vertically by controlling the water level in the excavated pit to enable the PRISM analysis technique: a mono-energetic 
%muon (anti) neutrino 
\num (\numb)
beam is synthesised from 
%measures 
the combination of measurements
of the energy spectrum at different off-axis angles, between $1.5^{\circ}$ and $4^{\circ}$ \cite{bhadra2014letterintentconstructnuprism,protocollaboration2018hyperkamiokandedesignreport}.
With a baseline of \SI{832}{\meter}, about $17,500$ CC \nue events in the neutrino-dominated beam mode and $19,000$ \nueb events in antineutrino-dominated beam mode will be detected.
The flotation compartments, the rail and the free bearing system are the key to controlling the detector position and precisely determining the off-axis angle. 
Although mechanical design work is ongoing, with its own unique challenges to realise the moving detector, facility construction work will begin in Summer 2025, starting from pit excavation. 
Multiple contributions from Europe are present in IWCD, and they encompass the detector tank and support structure, multi-PMTs, DAQ and calibration.

\begin{figure}[htb]
\centering
    \includegraphics[width=4.7cm]{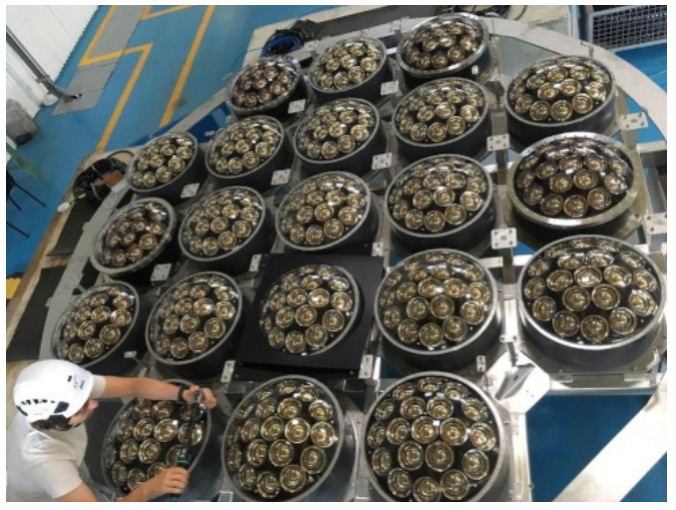}
    \includegraphics[width=4.7cm]{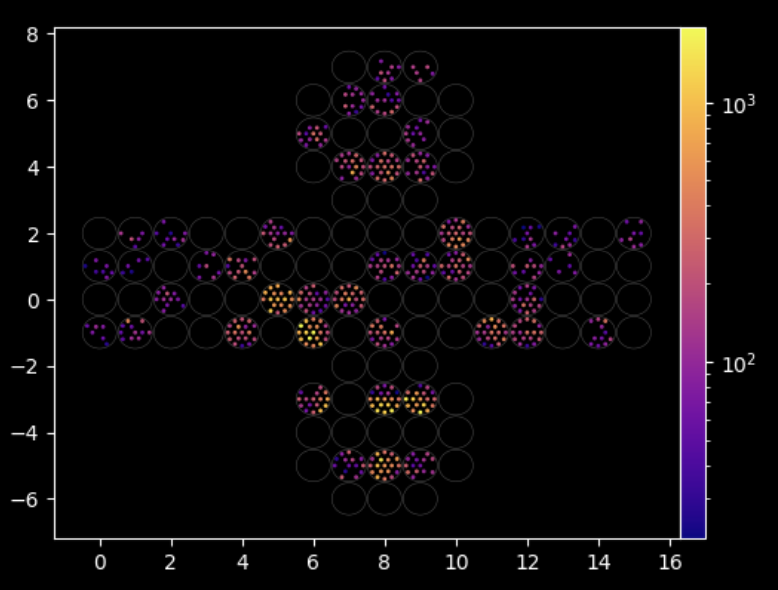}    
    \includegraphics[width=7.cm]{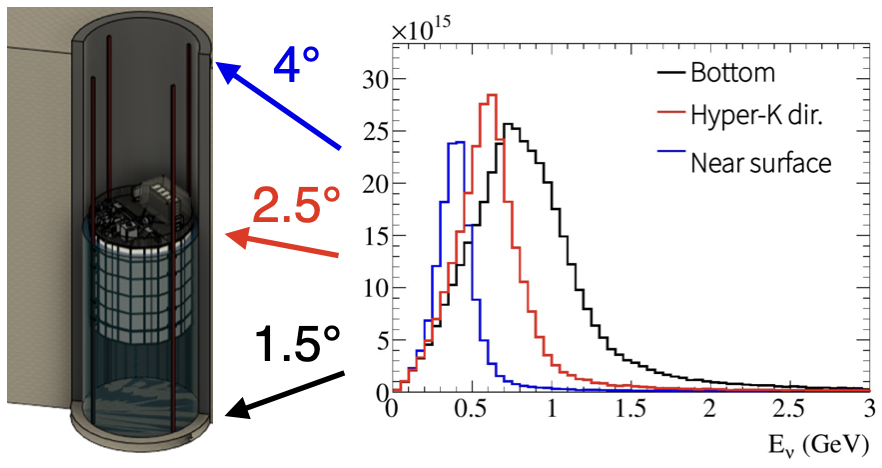} 
    \caption{
    Left: bottom part of the WCTE inner cylindrical tank instrumented with multi-PMTs. 
    Middle: event collected by WCTE during the 2024 data taking.  
    Top: representation of the IWCD system spanning the $1.5^{\circ}$ - $4^{\circ}$ off-axis angles and corresponding simulated neutrino energy spectra.
    }
    \label{fig:wcte}
    \label{fig:multi-pmt}
\end{figure}
%
%The \hkshort experiment will require unprecedented precision, hence 
The rejection of the \nue background in the \nue cross section measurement relies on the %electron-muon 
$e/\mu$
and 
%electron-gamma 
$e/\gamma$
separation,
that requires a detailed characterisation of the water-Cherenkov detector response.
%, including machine learning techniques.
% This is the goal of the ongoing Water Cherenkov Test Experiment (WCTE) at the CERN SPS \cite{Barbi:2712416}.
% With a test beam of different particles (muons, pions, electrons, neutrons and gammas) and momenta (200 - 1200 MeV/c),
% it deploys the multi-PMT technologies of both IWCD and the far detector, 
% %(4 units) 
% as shown in  Fig.~\ref{fig:wcte}).
%
%The Water Cherenkov Test Experiment (WCTE)
WCTE is taking data at the CERN SPS \cite{Barbi:2712416}.
Shown in Fig.~\ref{fig:wcte},
it deploys the multi-PMT technologies of both IWCD and the far detector to characterise the response of a water Cherenkov detector with known particle types (muons, pions, electrons, neutrons and gammas, between $200$ - $1200$\,MeV/c), 
and test and develop calibration systems.
%necessary for accurate modeling of the detector response and 
Moreover,
physics processes 
%including 
such as high-angle Cherenkov light production, pion scattering and absorption, and secondary neutron production in hadron scattering
can be studied.
%WCTE will be completed in 2026.
Although WCTE will be completed in 2026, \hkshort plans to use test beam facilities at CERN to characterize existing or new near detectors in the future.

\section{Physics programme}
\label{sec:physics}
\vspace{0.3cm}

The primary goal of \hkshort is the discovery of CP violation (CPV) in the leptonic sector and precision measurements of the neutrino oscillation parameters in the PMNS matrix. 

\textbf{CPV and mixing parameters: }
The \hkshort analysis is currently based on the well-established oscillation analyses performed by T2K, with modifications to take into account the novelties of the \hkshort design, including the upgraded T2K Near Detector and the addition of IWCD. 
 Due to the larger fiducial mass and the higher beam power, \hkshort will collect a thousands of selected events in both the \nue and \nueb appearance modes.

\noindent   Future analyses by \hkshort will be performed having access in the same experiment to both accelerator beam data and atmospherics neutrinos. 
This combination, as recently demonstrated by the joint analysis done by the T2K and \skshort collaborations~\cite{T2K:2024wfn}, will allow \hkshort to break the degeneracies between $\delta_{CP}$ and MO. Note that, as of today \skshort provides the most sensitive measurement of MO~\cite{Super-Kamiokande:2023ahc}.\\
\hkshort is expected to measure MO with a $3.8 \sigma~(6.2\sigma)$~C.L. for $\sin\theta_{23}=0.4~(0.6)$ after 10 years of data taking and, even in the unlikely case of no other detemination of MO is made, it would not significantly affect \hkshort abilities to discover CPV, as shown in Fig.~\ref{fig:hk_beamatm}.

\begin{figure}[!htb]
\centering
    \includegraphics[width=0.4\linewidth]{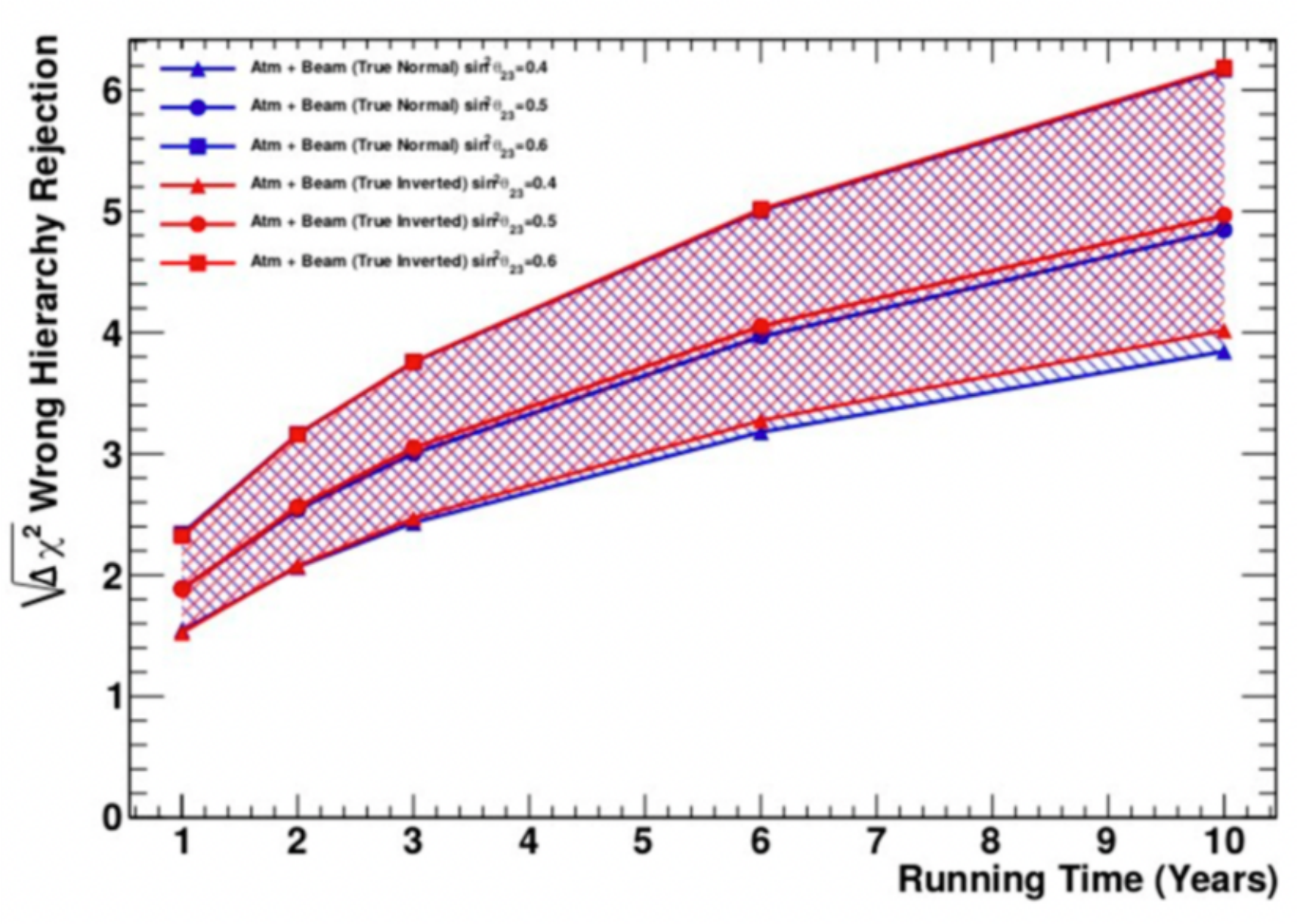}
    \includegraphics[width=0.45\linewidth]{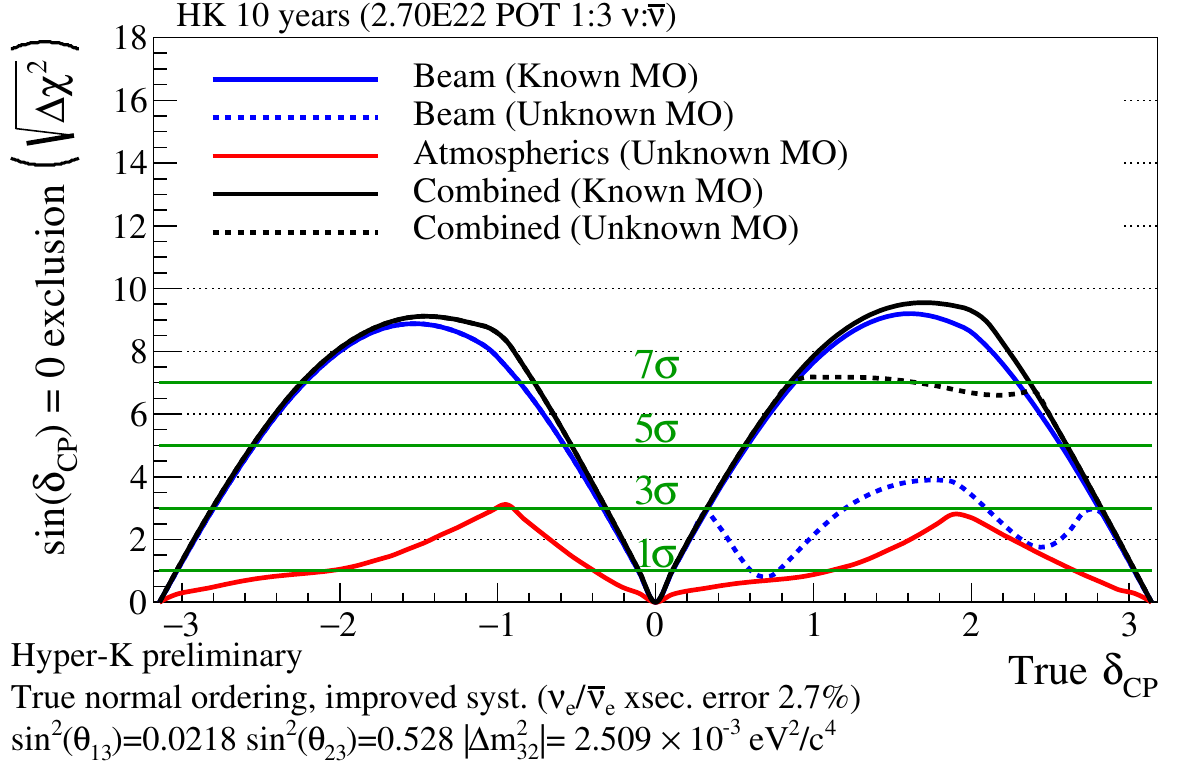}
    \caption{Left: Projected sensitivity for wrong mass hierarchy rejection by atmospheric neutrino analysis combined with accelerator neutrinos. For different $\theta_{23}$ octant assumptions; $\sin^ 2 \theta_{23}=0.4$ (triangle), $\sin^ 2 \theta_{23}=0.5$ (circle), and $\sin^ 2 \theta_ {23}=0.6$ (square). Blue and red colors indicate normal and inverted hierarchy for true cases.    
    Right: Sensitivity to exclude $\sin \left(\delta_{C P}\right)=0$, as a function of true $\delta_{C P}$ value, for $10$ \hkshort -years and true normal mass ordering.  
    The large sample of atmospheric neutrinos improves the sensitivity, particularly in the case of unknown mass ordering. The sensitivity for the opposite MO is analogous after swapping curves between positive and negative values of true \dcp.
    }
    \label{fig:hk_beamatm}
\end{figure}

\begin{figure}[!htb]
\centering
    \includegraphics[width=0.45\linewidth]{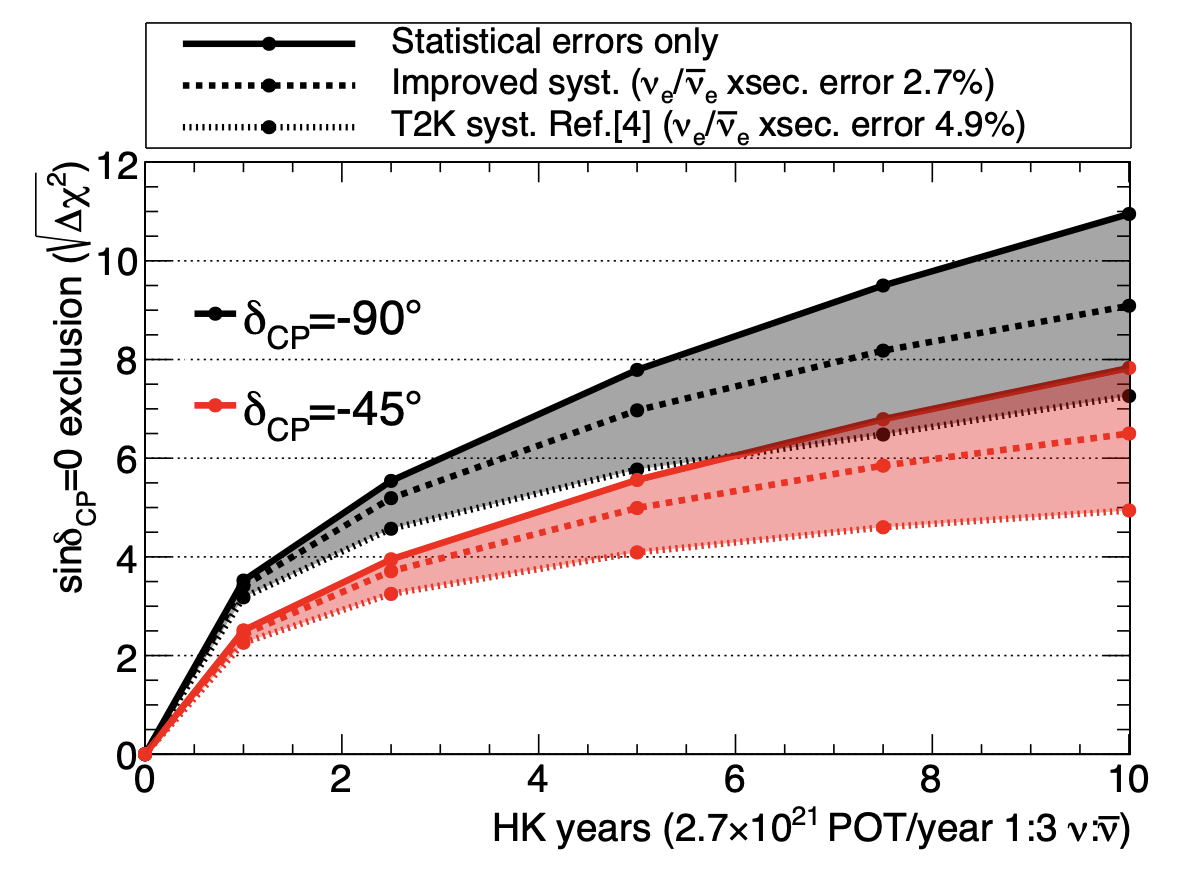}
     \includegraphics[width=0.45\linewidth]{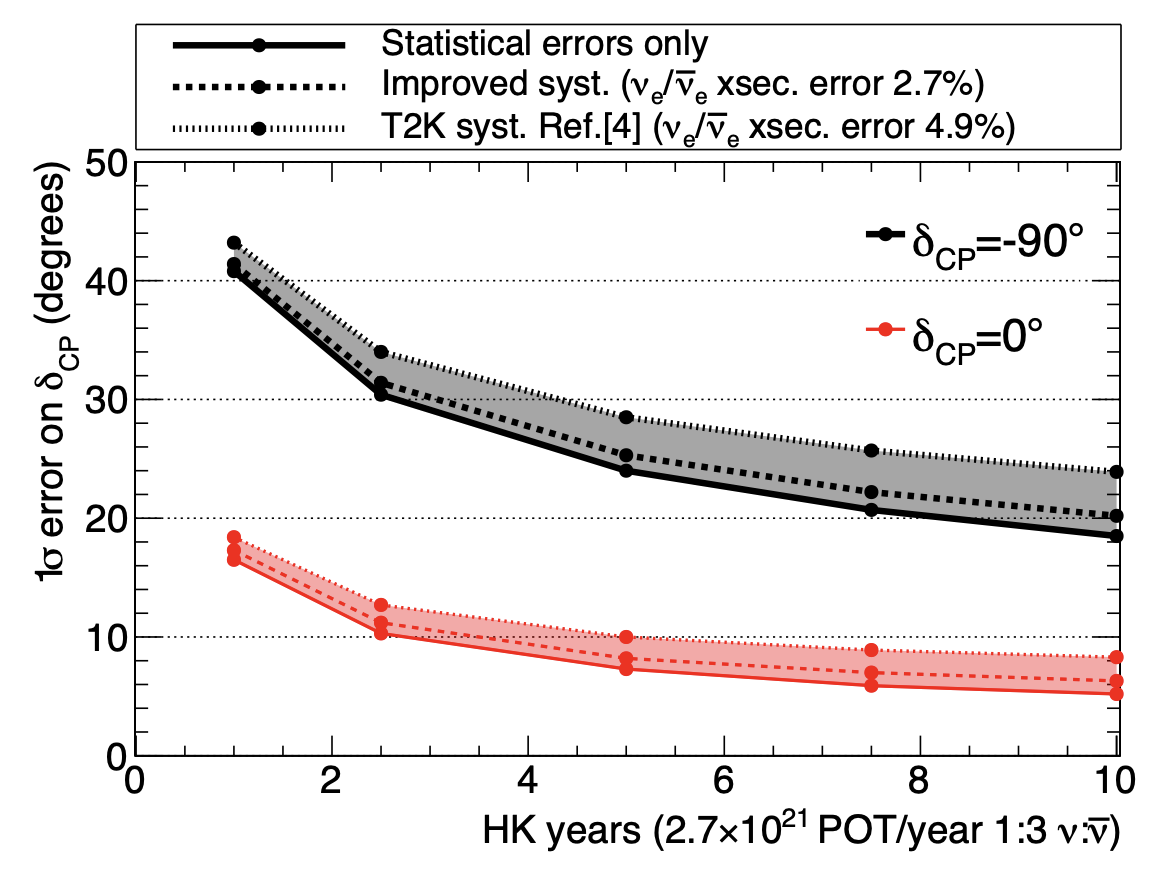}
       \caption{
Left: Sensitivity to exclude $\sin \left(\delta_{C P}\right)=0$ for true $\delta_{C P}=-\pi / 2$ and $\delta_{C P}=-\pi / 4$, as a function of \hkshort -years. The shaded areas show the span of possible values, for varying systematic error models.
    Right:     $1 \sigma$ resolution of \dcp as a function \hkshort years for $\dcp=-\pi / 2$ or 0.   
    }
    \label{fig:hk_nev}
\end{figure}
 
Focussing only on the accelerator beam to see its potential, we build a model of improved systematics by modifying the errors associated with each systematic parameter in the T2K model without modifying the correlations between the parameters. In particular, we highlight the improvement in the uncertainty in $\sigma\left(\nu_e\right) / \sigma\left(\bar{\nu}_e\right)$ which is a major systematic for \hkshort.
For comparison in the following plots, we also report sensitivities that use T2K uncertainties, as well as sensitivities that assume only statistical errors. 
These results highlight the impact that the systematic error model has on the physics reach of \hkshort.  Also, we assume that a neutrino to antineutrino beam running ratio of $1$ to $3$.%, to be decided at the time of data taking.

In the case of maximal CP violation in the leptonic sector, %as indicated by T2K measurements and 
as shown 
in Fig.~\ref{fig:hk_nev} (left), \hkshort will measure CP violation with more than $5\sigma$~C.L. after $3$-$4$ years of data taking tor $\dcp=-\pi/2$ and normal ordering, thanks to its large statistics and reduced systematic errors.

\noindent After 10 years of data taking, \hkshort will discover CPV at $>5\sigma$~C.L. for 60\% of the possible values of \dcp.
%. The exact ratio of neutrino versus antineutrino running will be optimised when \hkshort starts to take data. 
%\DS{Remove this last sentence ``The exact ratio...''. It's doesn't add anything}
The precision in measuring \dcp will range from $20^{\circ}$, in case of maximal violation CP, to $6^{\circ}$, for the CP conserving values, see Fig.~\ref{fig:hk_nev} (right).    
Beyond the search for CPV, \hkshort will also feature unprecedented precision on the atmospheric neutrino oscillation parameters, as shown in Figure~\ref{fig:hk_atmsector}.
\begin{figure}[!htb]
\centering
    \includegraphics[width=0.43\linewidth]{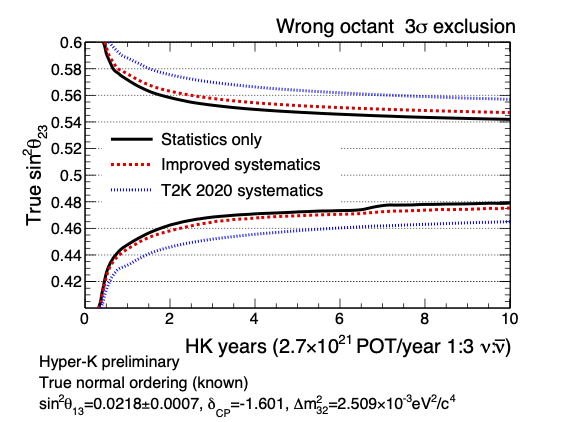}
    \includegraphics[width=0.43\linewidth]{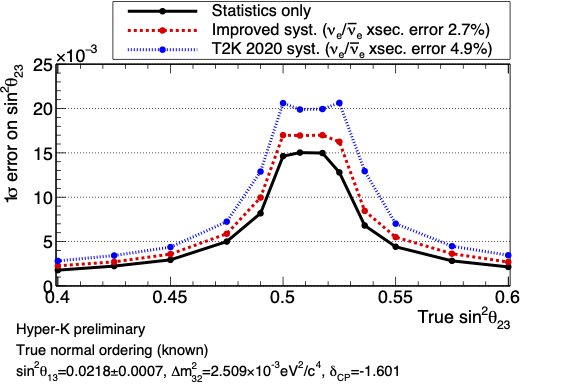}
 \caption{Left: Wrong octant $3 \sigma$ exclusion as a $2$D function of true $\sin ^2 \theta_{23}$ and \hkshort -years for $3$ different systematics models overlapped.
Right: $1\sigma$ resolution of $\sin ^2 \theta_{23}$ for $10$ \hkshort years as a function of the true value of $\sin ^2 \theta_{23}$.
    }
    \label{fig:hk_atmsector}
\end{figure}

Table~\ref{tab:oscresults} summarises the expected ultimate precision of the measurements of the oscillation parameters achievable by the \hkshort experiment.
\begin{table}[!htb]
    \centering
    \begin{tabular}{lll}
    \hline \hline Physics Target & Sensitivity & Conditions \\
    \hline \hline
    \multicolumn{3}{c}{Neutrino study w/ J-PARC $\nu$ - $1.3 \mathrm{MW} \times 10^8 \mathrm{sec}$} \\\hline\addlinespace[1mm]
    %5$\sigma$ CPV discovery if \dcp=$-\dfrac{\pi}{2}$ ($\dfrac{\pi}{4}$) & 3y (6 y) &  MO known (MO unknown) \\
    $5\sigma$ CPV discovery if \dcp=-$\dfrac{\pi}{2}$ (-$\dfrac{\pi}{4}$) & $3$y ($6$ y) &  MO known \\
    \dcp resolution ($1 \sigma $ )& $6.3^{\circ}$ ($20.2^{\circ}$) &$@$\dcp = $0, \pi$ (\dcp=$\dfrac{\pi}{2}$) \\
    $\sin ^2 \theta_{23}$ resolution & $0.5-3\%$ &  $@1\sigma$ \\
    $\Delta m^2_{23} $ resolution & $<0.5\%$ & $@1\sigma$\\ 
    $\sin ^2 \theta_{13}$ resolution & $<3\%$ &  $@1\sigma$, with reactor constraint \\
    $\theta_{23}$ octant determination & $[0,0.450] \cup[0.569,1]$ & $@5\sigma$ \\
    \hline \hline 
    \end{tabular}
    \caption{\hkshort mixing parameters discovery potential and resolutions. Value of the parameters fixed in the fit $\sin ^2 2 \theta_{12}=0.307$ , $\sin ^2 2 \theta_{13}=0.0218$, $\sin ^2 2 \theta_{23}=0.528$,  $\Delta \mathrm{m}^2 _{32} = 2.5$,    $\Delta \mathrm{~m}^2 _{21} = 7.53 10^{-5}$. Beam power $1.3$\,MW, $10$ years, $1:3$ neutrino:antineutrino ratio.}
    \label{tab:oscresults}
\end{table}

\textbf{Astrophysical Observatory: }
As an astrophysical observatory,
\hkshort will be able to measure many phenomena with high precision.
\hkshort will be able to observe the neutrino
flux from the next galactic core-collapse supernova in unprecedented detail. 
Once a galactic supernova happens, this ability will be a powerful tool for guiding simulations
towards a precise reproduction of the explosion mechanism observed in nature. \hkshort can detect neutrinos with energy down to \SI{3}{\mev} and
can point the supernova, due to its event-by-event directional sensitivity.  
\noindent \hkshort will observe the antineutrino emitted in the explosion of supernovas through the inverse $\beta$-decay and to neutrinos through their scattering with electrons~\cite{Hyper-Kamiokande:2021frf}. This latter channel also provides strong directional information, since electrons are mostly scattered in the forward direction. The number of expected events as a function of the distance of the supernova explosions are shown in Fig.~\ref{fig:hk_sn}, left. \hkshort will have the unprecedented ability to detect neutrinos from supernovae beyond the Milky Way: for a supernova explosion in the Large Magellanic Cloud, at $50$\,kpc distance, \hkshort would detect about $3000$ events, while for a supernova in the Andromeda galaxy (M31), at $780$\,kpc distance, $\mathcal{O}(10)$ events are expected. 
\hkshort's ability to discriminate between the core collapse mechanism models was published in Ref.~
\cite{Hyper-Kamiokande:2021frf}.

\begin{figure}[htb]
\centering
    \includegraphics[width=0.37\linewidth]{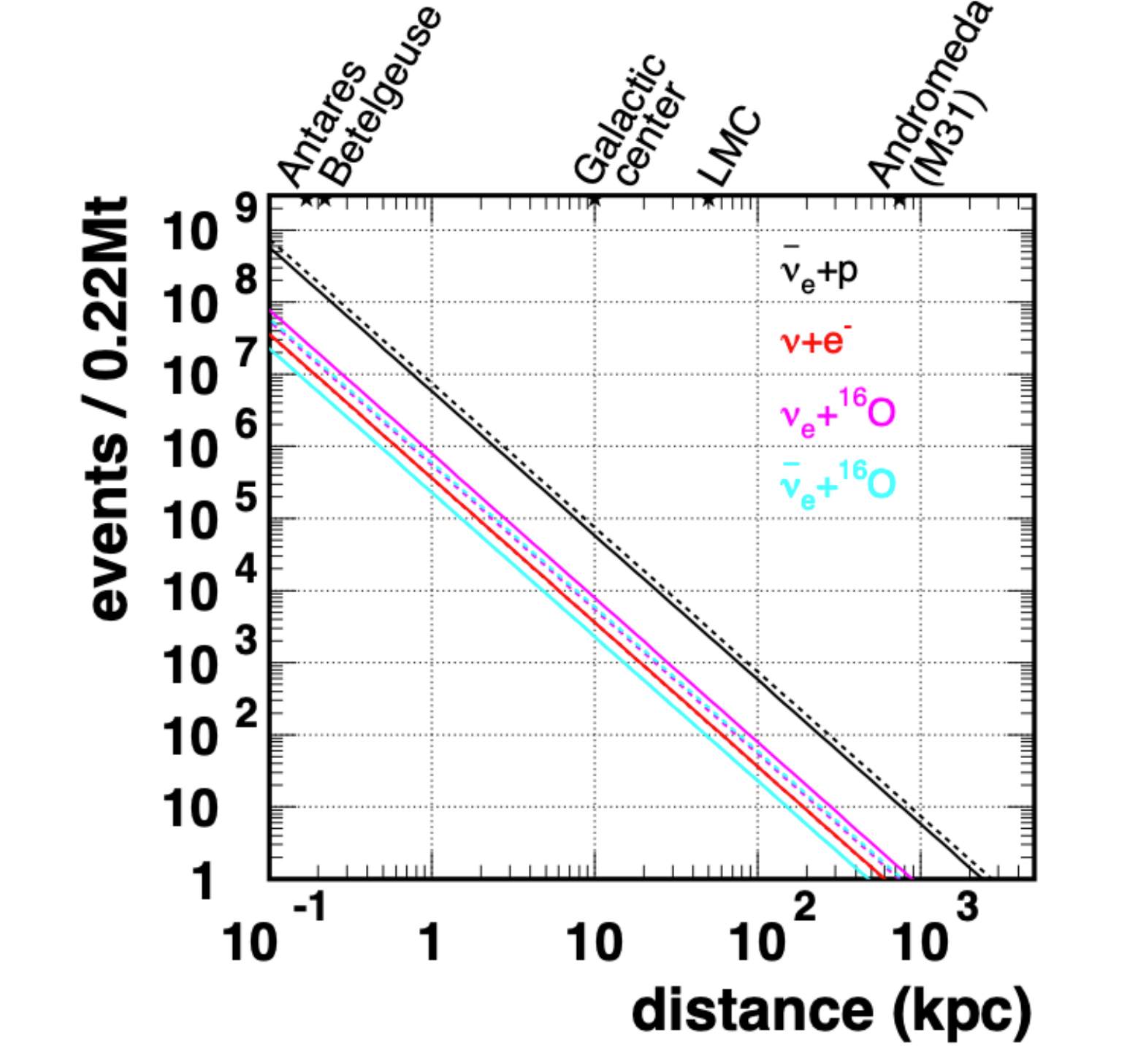}
    \includegraphics[width=0.45\linewidth]{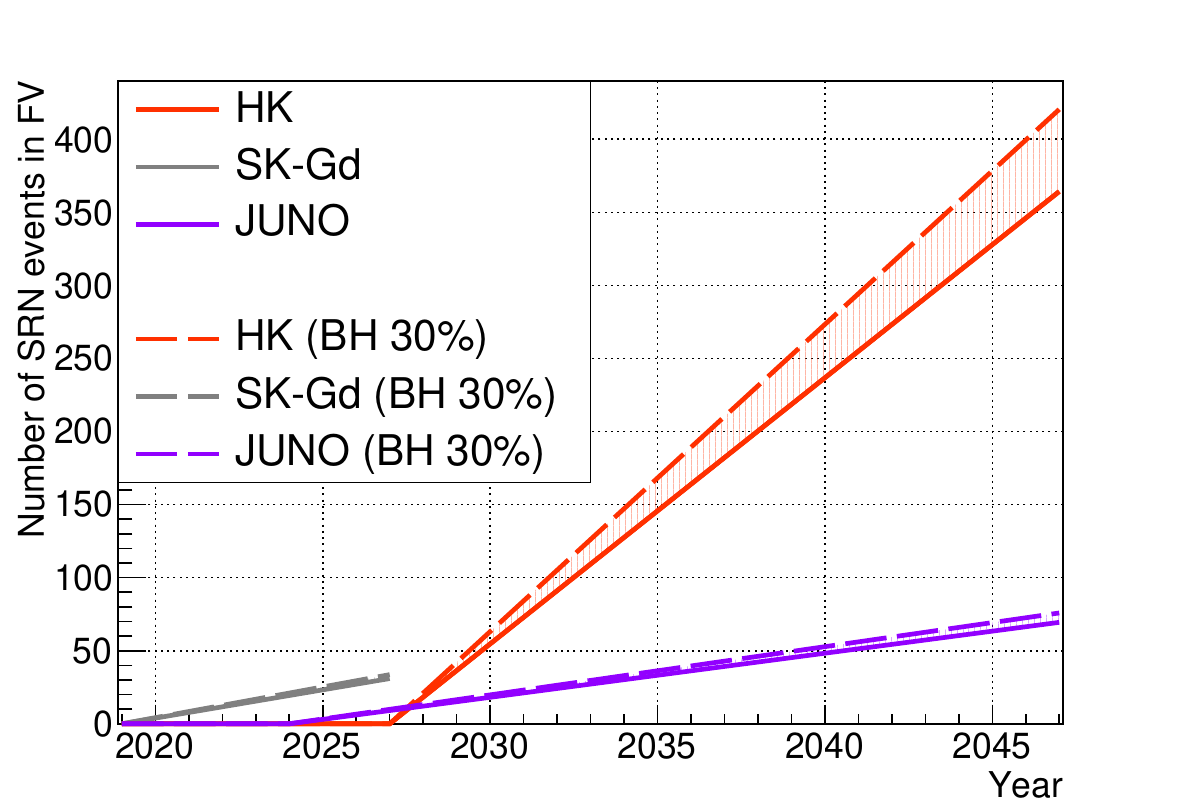}
    
    \caption{
    %\textcolor{red}{(Improve plot resolution)}
    Left: Expected number of events as a function of the distance of the SN explosion in the \hkshort volume. Right: Expected number of DSNB interactions in the \hkshort fiducial volume as a function of the years compared with \skshort and JUNO.
    }
    \label{fig:hk_sn}
\end{figure}

In addition, \hkshort will be sensitive to the diffuse supernova neutrino background (DSNB), detecting neutrino interactions emitted by SN explosions throughout the entire history of the Universe. The expected number of events as a function of the year is  shown in Fig.~\ref{fig:hk_sn}, right. 
\hkshort aims to discover and characterise the diffuse SN flux owing to a high statistics sample of antineutrino inverse beta decay interactions.

Solar neutrino measurements can achieve good precision with respect to the $\Delta m_{21}^2$ and day/night asymmetry of solar $\nu$ flux caused by terrestrial matter effects (as indicated by \skshort \cite{Super-Kamiokande:2023jbt}). 
These measurements could explain the variation of the solar neutrino flux and the "upturn" of the solar neutrino spectrum; furthermore, they may mitigate the tension between the $1$-$2$ mass splitting extracted from the global fit of the solar neutrino data $\Delta m_{21}^2=\left(4.7_{-0.17}^{+1.6}\right) \times 10^{-5} \mathrm{eV}^2$ and the value measured by KamLAND with $\Delta m_{21}^2=\left(7.5_{-0.17}^{+0.19}\right) \times 10^{-5} \mathrm{eV}^2$. 

Through the observation of $\sim$\SI{10}{\mev} neutrinos with time, energy, and directional information, \hkshort will take a unique role as a multi-messenger observatory, expanding on the successful multi-messenger science programme of \skshort. It has the potential to detect thermal neutrinos from nearby ($<10~\mathrm{Mpc}$) neutron star merger events in coincidence with gravitational waves.

\hkshort can also investigate a variety of other astrophysical phenomena, indicating its wide potential as observatory. 
For example, transitional events such as solar flares will give us important information about the mechanism of the particle acceleration at work. 

\hkshort also has the potential to see neutrinos from astrophysical sources such as magnetars, pulsar wind nebulae, active galactic nuclei, and gamma-ray bursts.

\textbf{Nucleon decays: } 
Optimising \hkshort for the observation and discovery of a nucleon decay signal is one of its main design drivers.
Due to its huge mass, and thus number of protons, \hkshort will be able to significantly extend the sensitivity beyond existing limits, many of which have been established by \skshort.
\hkshort will have unprecedented sensitivity to proton decay, surpassing existing limits from \skshort in the $p \rightarrow e^+\pi^0$ channel 
and leading searches in the $p \rightarrow \nu K^+$ channel, as shown in Fig.~\ref{fig:hk_pdecay}.

\begin{figure}[htb]
\centering
    \includegraphics[width=0.45\linewidth]{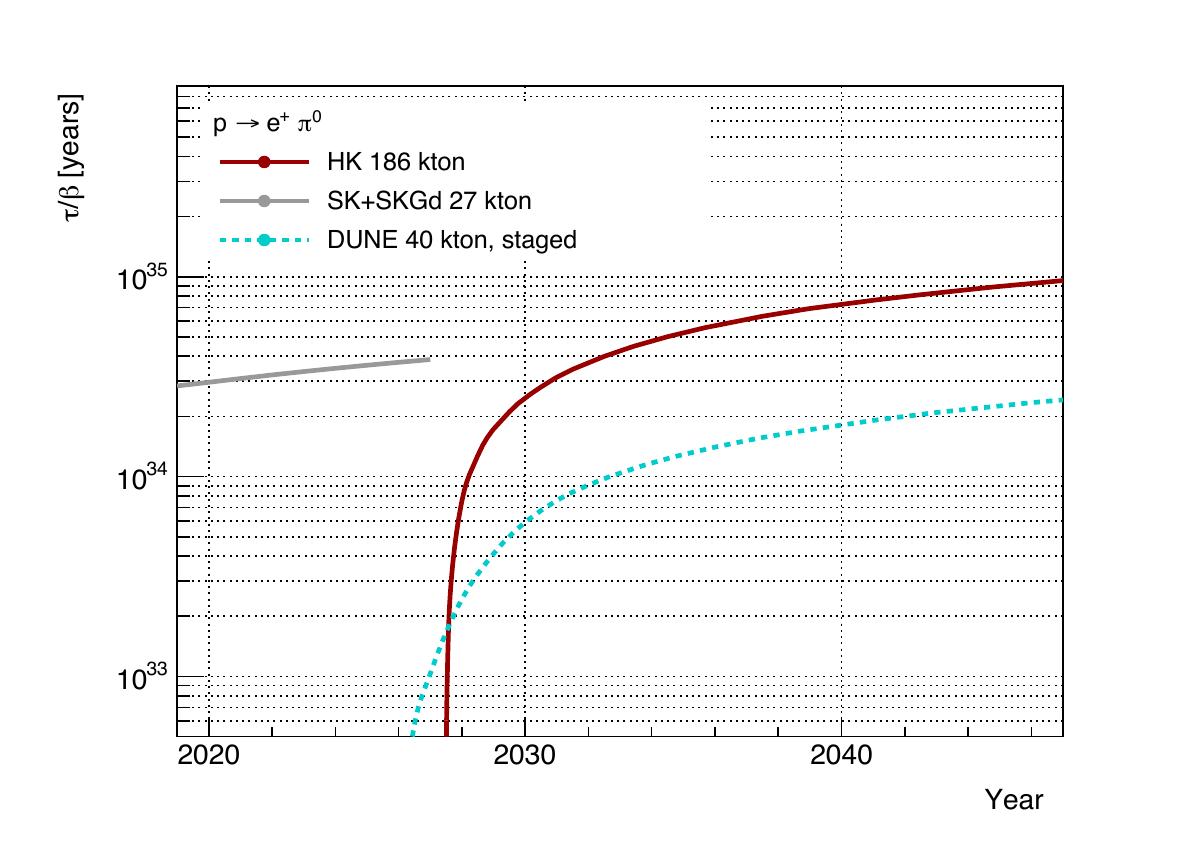}
    \includegraphics[width=0.45\linewidth]{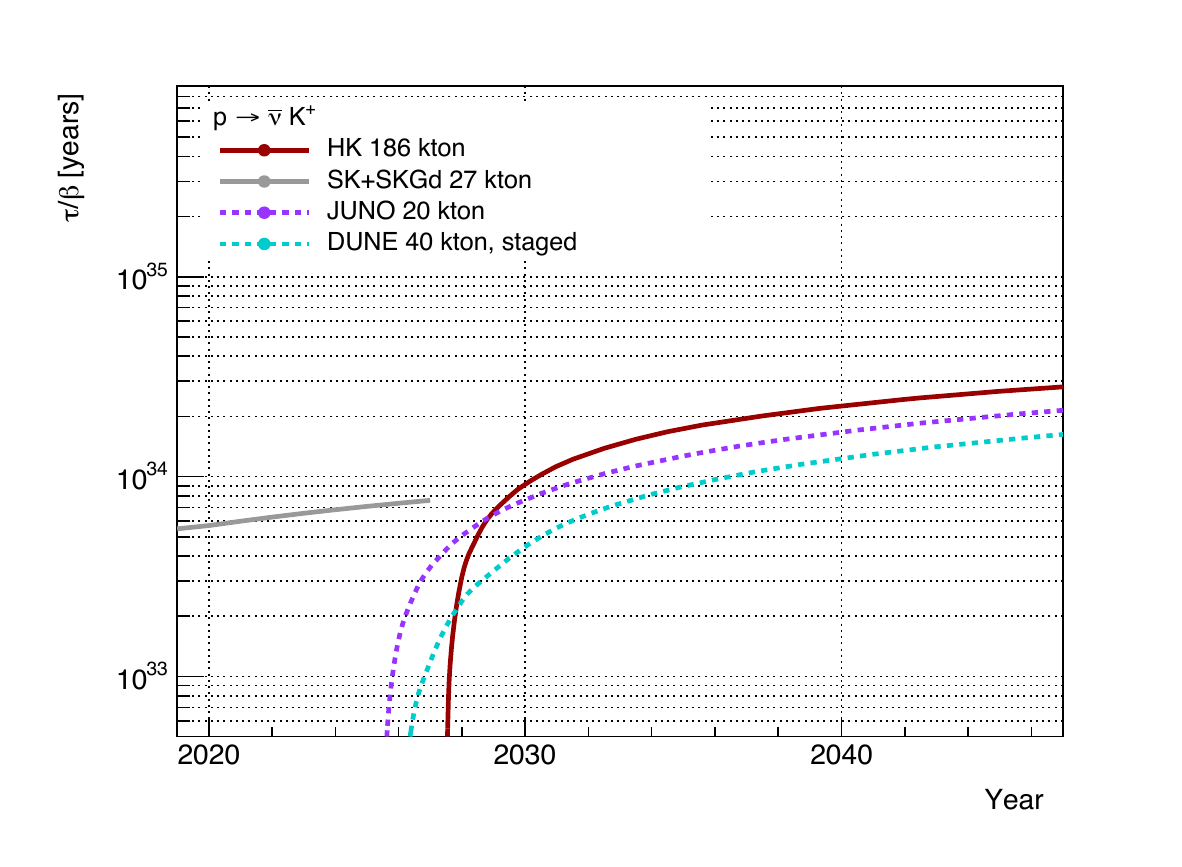}
    \caption{\hkshort predicted sensitivity to the  $p \rightarrow e^+\pi^0$ and to the $p \rightarrow \nu K^+$ channels for proton decay compared with \skshort, DUNE and JUNO. }
    \label{fig:hk_pdecay}
\end{figure}

Although the decay mode $p \rightarrow e^{+} \pi^0$ is predicted to be dominant in many GUT models, a variety of other decay modes are possible with different mesons and leptons, each with a sizable branching ratio (see Ref.~\cite{protocollaboration2018hyperkamiokandedesignreport}) and accessible to \hkshort. The diversity in these predictions suggests that to make a discovery and subsequently constrain proton decay models,
it is critical to probe as many nucleon decay modes as possible, as different branching fractions can help to discriminate between the models.

\textbf{Dark Matter and Beyond the Standard Model Physics Searches:} 
\hkshort can search for dark-matter WIMPs by looking for neutrinos created in pair annihilation from trapped dark matter in the Galactic centre or the centre of the Sun. 
\hkshort will have the ability to detect both the $\nu_e$ and $\nu_\mu$ components of the signal, making it more sensitive to this type of analysis \cite{Super-Kamiokande:2022ncz}.

The \hkshort Near Detector complex will also enable searches for BSM processes such as heavy neutral leptons (HNL), Lorentz and CPT symmetries. Previous studies are reported e.g. in Ref.~\cite{T2K:2017ega}, but they will benefit from the higher intensity beam.
%and sterile neutrinos. 
T2K has already performed a search for HNL in the ND280-OA TPC gas~\cite{T2K:2019jwa}. Profiting from the higher beam power and the additional TPCs installed with the upgrade of ND280-OA, \hkshort will be able to improve these measurements. 
%Searches for a class of extensions of the Standard Model that violate the Lorentz and CPT symmetries have been searched with the INGRID detector \cite{T2K:2017ega} and will benefit of the higher intensity beam.
The new fine grained target mass and performance with the upgrade of ND280-OA improves the sensitivity to short-baseline oscillations due to sterile neutrinos, whose first search has already been performed at T2K \cite{T2K:2014xvp}. 
The combination of ND280 (\SI{280}{m} distance from the target) and IWCD (at a distance of 850~m) will allow powerful $2$-detector searches for sterile neutrinos with $\Delta m^2 \sim 1~eV^2$.

\section{Program for the ultimate reduction of systematic uncertainties}
\label{sec:ultimate-syst-unc}

\subsection{The \hkshort high-statistics phase}
\label{sec:high-stat-phase}

\vspace{0.3cm}

After five years since the start of data-taking, the systematic uncertainty will become dominant.
In Fig.~\ref{fig:cpv-vs-time-vs-syst} one can see that, by reducing the \sigmanuenueb systematic uncertainty, the fraction of truth \dcp values s=for which \hkshort can exclude the CP conserving hypothesis increases more than by running for a longer time.
Current sensitivity studies show that 
ND280 and IWCD, together, constrain
the \sigmanuenueb systematic uncertainty down to $4$\%.
If reduced down to $2.7$\%, CP violation can be discovered for almost $75$\% of truth \dcp values after $10$ years of data taking.

Thus, the \hkshort collaboration is developing the conceptual design of the final upgrade of ND280-OA, called ND280++, envisaged for the high-statistics phase, after 2030.
ND280++ would complement the upgrade of 2023 by replacing the sub-detectors that had been installed in 2009, 
i.e. the two FGDs and the three vertical TPCs, %that have been running since the start of T2K data-taking in 2009, 
with $10$ tonnes of water and/or organic scintillator detectors, to obtain a neutrino target mass three times higher than the current one. 
%As ND280 is much closer than IWCD to the neutrino production target, 
ND280++ could almost double the overall \hkshort ND280-OA neutrino event rate.
%\DS{a sentence about the expected number of events ?} \CG{I do not understand this sentence.}
%
The goals of ND280++ are 
(a) the precision measurement of the \num and \numb interaction cross section in water, 
(b) high-resolution reconstruction of the hadronic final state, 
(c) the collection of a high-statistics sample of \nue and \nueb interactions.
The reference design for ND280++, shown in Fig.~\ref{fig:nd++}, includes
$4.5$ tonnes of inactive water alternated with tracking planes of scintillating fibres;
$4.5$ tonnes of $3$D segmented scintillating water with a design analogous to SuperFGD but with at least $80$\% by mass of water;
(3) a $1$ tonne scintillating fibre (scifi) target 
%that alternates X- and Y-oriented mats 
to track protons down to $150$\,MeV/c;
(4) a new vertical TPC.
%thinner and more compact than those already in ND280. 
%
The ND280++ alternative design includes a multitonne SuperFGD-like plastic scintillator detector. 
Simulation studies and R\&D are ongoing at various European institutes. 
The development of new detector technologies involves: the
water-based liquid scintillator \cite{Yeh:2011zz} organised in a $3$D matrix of optically-isolated \SI{1}{\cm}$^3$ voxels (WbLS-$3$D);
%with highly-reflecting materials integrated into a 3D-matrix of 1~cm$^3$ voxels \cnn{paper in preparation}, 
the integration of $2 \times 2~\text{m}^2$ scifi planes; 
the cost-effective readout of submillimetre diameter scintillating fibres with monolithic single-photon avalanche diode (SPAD) arrays \cite{Franks:2023ttv,Buhrer:2024sio};
%imaging monolithic sensors for a  of the scintillation light \cnn{spad-scifi}
%are being developed.
$3$D printing of segmented plastic scintillator \cite{Berns:2020ehg,3DET:2022dkw,Weber:2023joy}; precise gluing of plastic scintillator cubes \cite{Boyarintsev:2021uyw} and a segmented cast scintillator.
%\DS{should we include also high-p TPC ?}
Part of this R\&D is also conducted within the CERN Detector R\&D (DRD) collaborations \cite{Colaleo:2885937,CERN-DRDC-2024-001} and the $3$DET R\&D collaboration hosted by CERN \cite{3det}.

This project will give a unique opportunity to develop, test and deploy novel technologies with real large-size implementation.
Building on the success of NP07 with the first upgrade of ND280, the support of CERN to ND280++ with detector prototyping, tests and assembly will be crucial.
\begin{figure}[htb]
\centering
    \includegraphics[width=8.5cm,height=5.2cm]{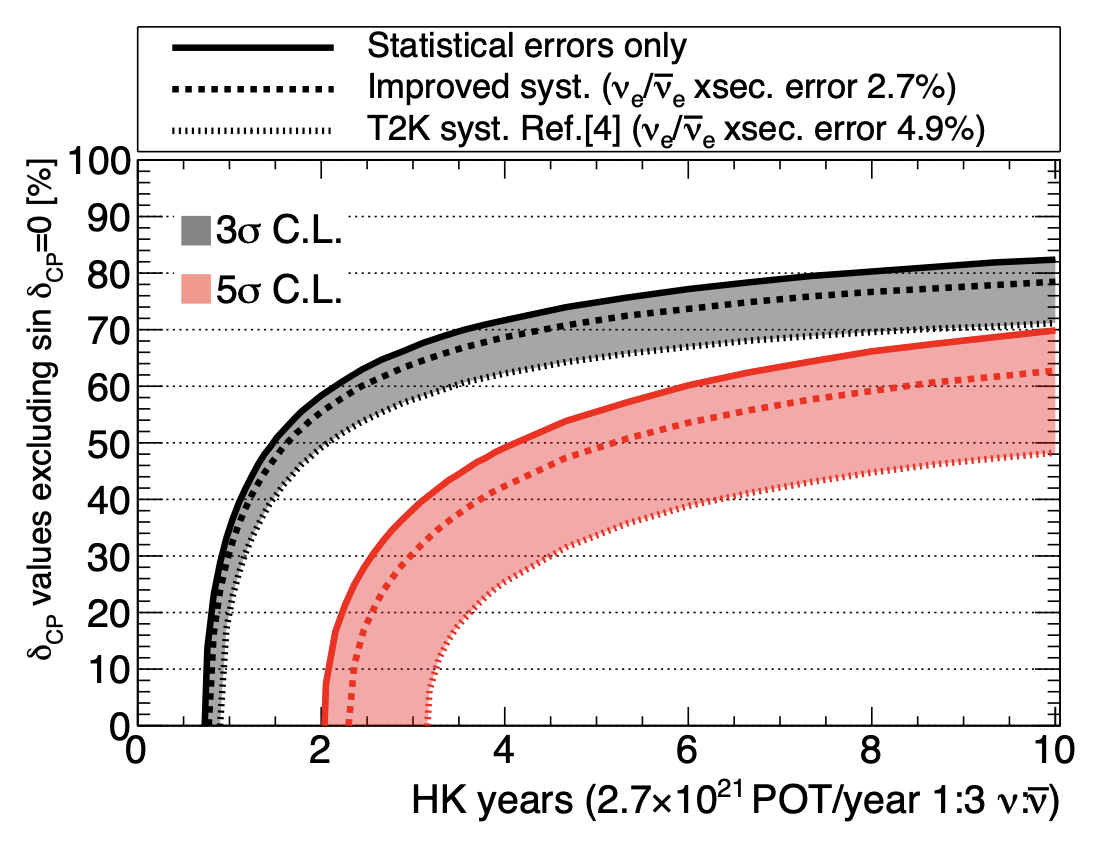}   
    \includegraphics[width=8.cm]{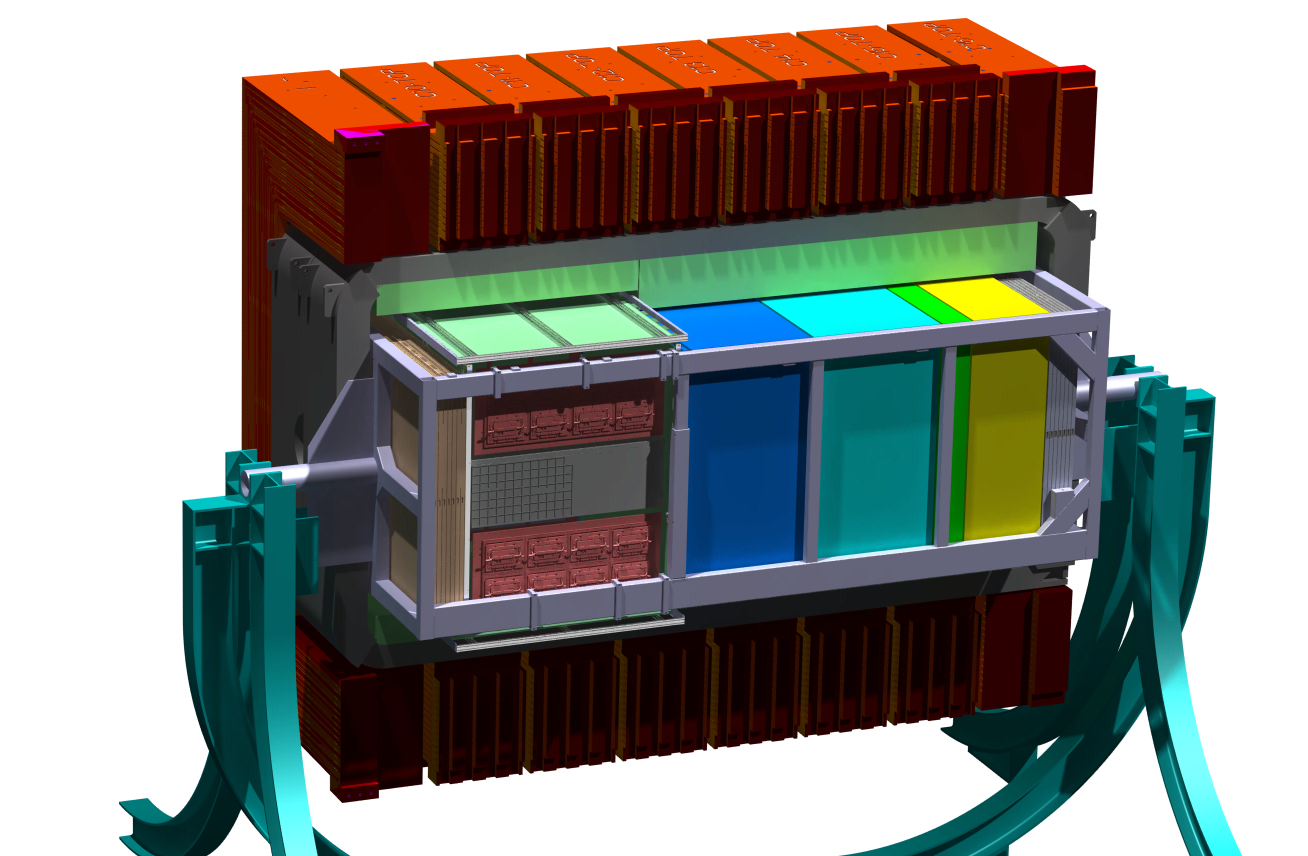}
    \caption{
    Left: fraction of truth \dcp values for which the CP conserving hypothesis ($\sin \dcp = 0$) is excluded at the $3\sigma$ (black) and $5\sigma$ (red) confidence level as a function of the \hkshort running time, for three different systematic error assumptions. 
    Right: cross section of the ND280++ reference design. The detectors are enclosed within the UA1 magnet. The upgrade would affect the following volumes: inactive water detector (dark blue), WbLS-3D detector (light blue), the scifi (green) and a new vertical TPC (yellow).
    }
    \label{fig:cpv-vs-time-vs-syst}
    \label{fig:nd++}
\end{figure}
%\subsection{Auxiliary hadron-production measurements for the neutrino flux prediction}
\subsection{Auxiliary measurements for \hkshort at CERN}
\label{sec:flux-constraint}

\textbf{Hadron-production measurements at NA61/SHINE:}
the accurate prediction of the \hkshort neutrino flux primarily relies on the hadron-production measurements performed at the auxiliary \nashine experiment at CERN SPS \cite{NA61:2014lfx}.
Protons are accelerated up to \SI{30}{\gev} and  directed onto a graphite target to reproduce the configuration of the J-PARC neutrino beamline and measure the decay products of (mainly) charged pions and kaons.
Previously, the T2K replica target (\SI{90}{\cm} long) allowed to constrain the prediction of the neutrino flux, affected by the modeling of secondary meson interactions, from an uncertainty of $10$\% down to $5$\% at the flux peak~\cite{NA61SHINE:2016nlf,NA61SHINE:2018rhe}.
With a \SI{2}{\cm} thin target, measurements of the pure proton-carbon cross section were performed~\cite{NA61SHINE:2015bad}.
\nashine is an important asset for the European community and played a crucial role in ensuring the success of T2K. Improved measurements of hadron-production will be extremely beneficial to \hkshort.
We plan to make new measurements with the \hkshort replica target \cite{Aduszkiewicz:2309890}, including new designs to reduce the wrong-sign antineutrino (neutrino) contamination.

\textbf{Low energy beam for \nashine:}
the \nashine hadron-production measurements were intended to support discovery of electron neutrino appearance and require an additional improvement to fully constrain the \hkshort neutrino beam flux and atmospheric flux.
For instance, a low uncertainty on the ``wrong-sign'' component in the neutrino flux, 
i.e. the contamination of neutrinos (antineutrinos) in a antineutrino (neutrino) beam,
is vital for the search for CP violation.
%The proposal is to build a 
Thus, the implementation of a new branch of the H2 beamline at CERN to also deliver low-energy ($2$-$13$ GeV/c) hadrons to NA61/SHINE has been proposed~\cite{Nagai:2810696} and it would be very beneficial for \hkshort (Fig.~\ref{fig:low-energy-beam}, left).
Furthermore, these new data will also constrain the atmospheric neutrino flux. They can significantly reduce the flux uncertainty (Fig.~\ref{fig:low-energy-beam}, right) which reduces the uncertainty on the atmospheric neutrino flux for the \dcp parameter.
The goal is to realise the new beamline at CERN and start a new measurement campaign right after the LHC long shutdown 3.
\begin{figure}[htb]
\centering
    \includegraphics[width=6.5cm]{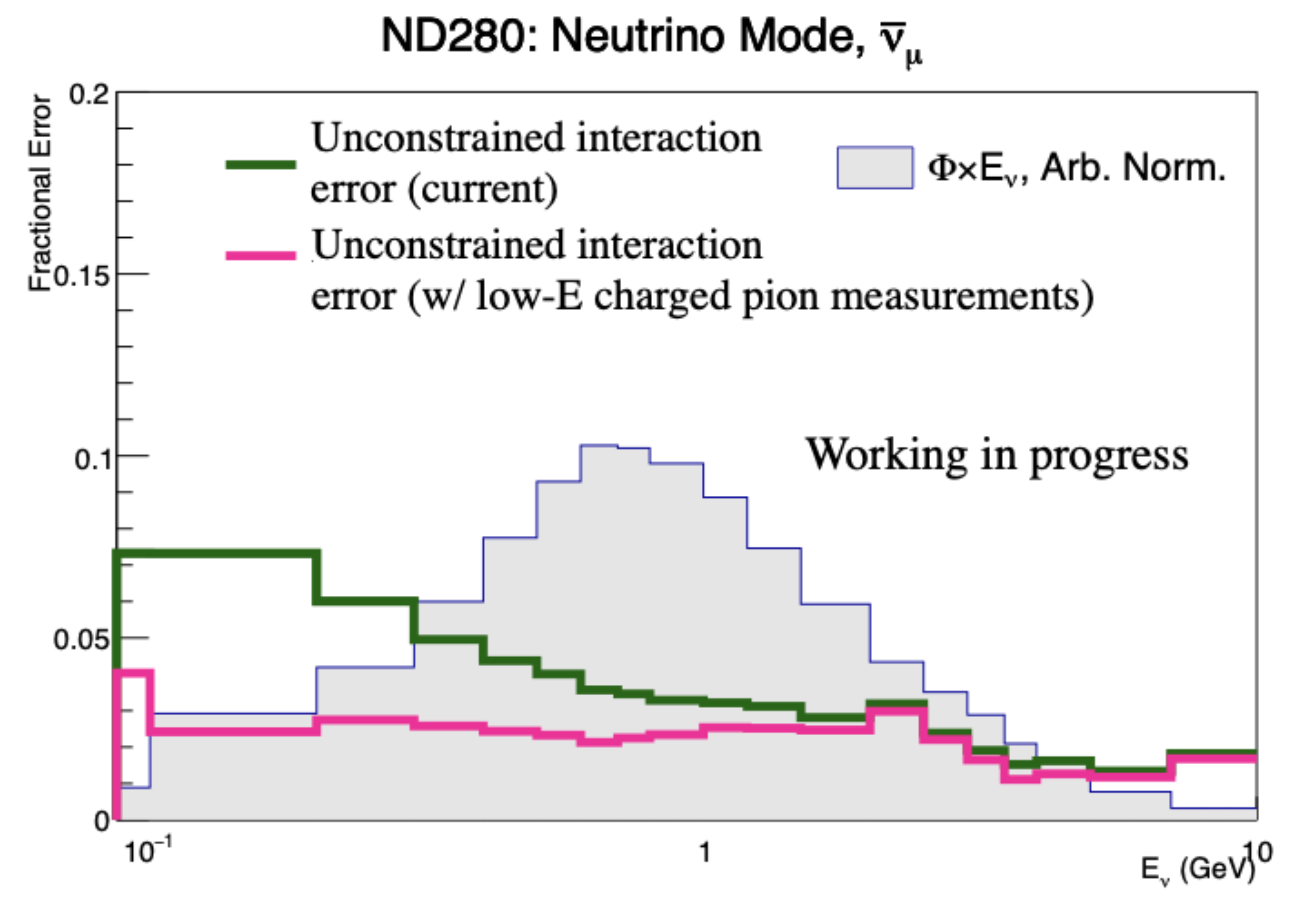}
    \includegraphics[width=6.5cm]{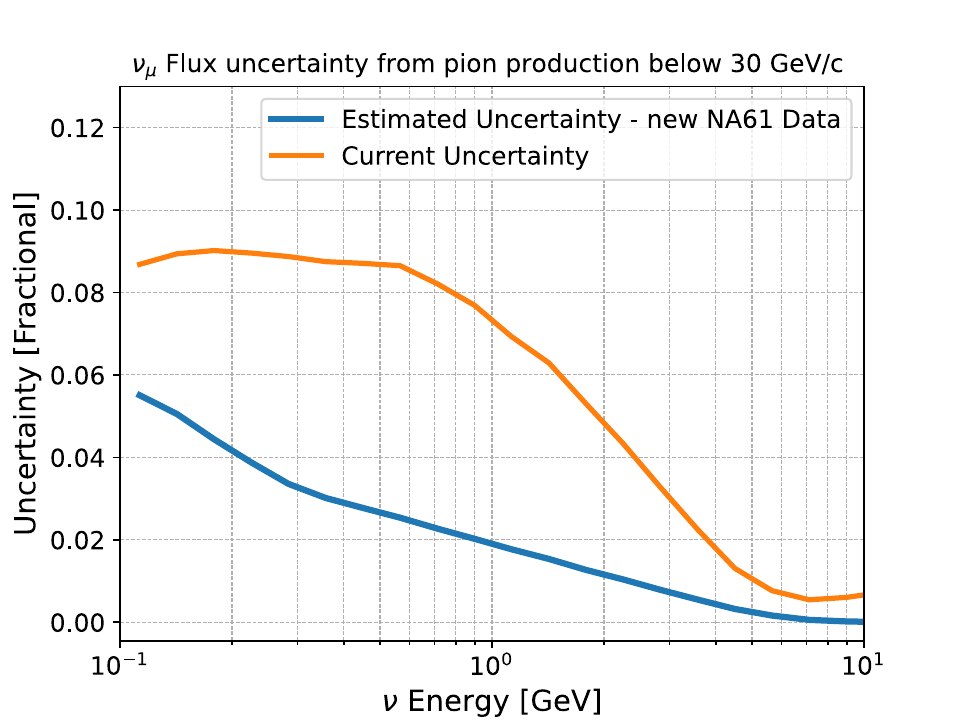}
    \caption{
%    Left: CAD drawing of the low-energy branch of the H2 line at CERN, starting at the secondary target and ending at the \nashine TPC \cite{Nagai:2810696}.
    Left: expected improvement on the unconstrained interactions in the \hkshort ND280 flux under the assumption of charged pion interaction measurements at the low-energy hadron beamline with a precision similar to the already performed pC$@$31 GeV/c measurements \cite{Nagai:2810696}.
    Right: Atmospheric neutrino flux uncertainties come from the $p+\mathrm{N} \rightarrow \pi^{ \pm}+\mathrm{X}$ process for muon neutrinos. Only the uncertainties due to $<30$\,GeV hadronic interactions are considered. Each plot shows uncertainty with and without new low-E NA61/SHINE measurements.
    Figures are from Ref.~\cite{Nagai:2810696}.}
    \label{fig:na61}
    \label{fig:low-energy-beam}
\end{figure}

An MoU is in place between \hkshort and \nashine for the common projects described above.

\section{Outlook}
\label{sec:outlook}
\textbf{Prospects for the next five years:}
\hkshort will have the best sensitivity to the discovery of the leptonic CP violation and will be able to determine the neutrino mass ordering with significance between $4$-$6$ standard deviations on its own by combining the accelerator and atmospheric high-statistic neutrino data.
As a multi-purpose observatory, supernova burst and background neutrinos as well as solar neutrinos can be efficiently detected. 
Due to its large mass, \hkshort will have a unique sensitivity to the detection of proton decay.
Finally, searches for 
%feebly interacting particles 
dark matter candidate particles
can be performed both at the near and the far detectors.
In the next five years, \hkshort will complete the construction of the massive water Cherenkov far detector,
the neutrino beam will be delivered by J-PARC at its maximum intensity of $1.3$\,MW,
and both IWCD and ND280 will be fully operational.
\hkshort will collect about two years of physics data, and the first results on the search for CP violation will be reported. 
%The contribution of the European particle physics community is 
If $\sin \dcp$ is close to $\pm \pi/2$, \hkshort could observe a significant hint of leptonic CP violation for its first results.  
The continuous support of the European particle physics community to the detector construction, operation, and, eventually, to the data exploitation is of vital importance to ensure the realisation of such programme.

\textbf{Ultimate reduction of systematic uncertainties:}
A detailed programme is being implemented by \hkshort to finalise systematic uncertainty.
Further reducing the neutrino flux constraint will require new auxiliary hadron-production measurements at the \nashine experiment at CERN, including the realisation of the low-energy hadron beam.  
\hkshort needs to realise the new beamline and start new hadron production measurements immediately after the LHC Long Shutdown $3$.

\noindent A major source of systematic uncertainty comes from the modelling of the neutrino-nucleus cross sections.
Currently, the \hkshort collaboration is developing the conceptual design of ND280++, the final upgrade of ND280, for the high-statistics phase, after 2030, towards the ultimate precision in the measurement of the CP violating phase.
The implementation of the ND280++ project should build on the experience achieved with the successful first upgrade of ND280 and NP07. 
Thus, it is crucial for \hkshort to receive the necessary support from the CERN Neutrino Platform for detector prototyping, tests, and final assembly. 
ND280++ is also a unique opportunity for the high-energy physics community to boost the development of new technologies by implementing and operating them at large-scale.
The environment provided by the DRD collaborations at CERN will play an important role in the completion of the detector R\&D.

\noindent Although \hkshort will constrain these uncertainties with near detectors, additional worldwide efforts are being made for neutrino cross-section measurements and
%\DS{I don't understand what is meant here. Where would these xsec measurements happen ? SBN$@$CERN ? Other ?}
the development of accurate nuclear and theoretical methods to compute neutrino cross sections. These efforts would contribute to maximise the results of the \hkshort experiment and further improve its unique sensitivity to CP violation. 
In this context, if a neutrino beam is built at CERN, it should also be designed to produce neutrinos in the energy range of interest for \hkshort.  

\textbf{Knowledge exchange:}
\noindent The work in \hkshort is characterised by a strong industrial return that will continue until the construction of the experiment and with future upgrades to further reduce systematic errors in the high-statistics phase. Close association with industrial partners promotes knowledge exchange. 
It will constitute a unique and exciting environment for training and forming the new generation of particle physicists. 
The \hkshort time schedule will allow the development of skills in any aspect of a large-scale particle physics experiment, such as detector R\&D, construction, operation and data analysis.
More widely, the trained workforce within the experiment will have developed skills for both industry and academia.  
We expect hundreds of papers will be published by \hkshort, accompanied by related seminars and conference talks. Outreach events to disseminate the results of the experiment will be organised.
%\DS{IMO, that we give talks and seminars is relevant for a grant request but for this type of document. It's obvious...}

\begin{tcolorbox}[colback=lime!10!white]

\textbf{Final considerations:} In summary, \hkshort will address the biggest unsolved questions in physics leaving its footprint in the future of particle physics through a multi-decade physics programme that will start in 2027 and will run for at least 20 years. It is important for Europe to continue and strengthen its involvement and leadership in the experiment towards a successful future.

\end{tcolorbox}

\section*{Acknowledgements}

The Hyper-Kamiokande Collaboration would like to thank
the Japanese Ministry of Education, Culture, Sports, Science and Technology (MEXT), the University of Tokyo, Japan Society for the Promotion of Science (JSPS), the Kamioka Mining and Smelting Company, Japan; 
% Armenia
the Minister of Education, Science, Culture and Sports, grant 21T-1C333, Armenia;
% Australia
%, Australia;
% Brazil
CNPq and CAPES, Brazil;
% Canada
%, Canada;
% Czech Republic
Ministry of Education Youth and Sports of the Czech Republic 
(FORTE {CZ.02.01.01/00/22\_008/0004632}); %, Czech Republic;
% France
CEA and CNRS/IN2P3, France;
% Germany
%, Germany;
% Greece
%, Greece;
% India
%, India;
% Italy
the INFN, Italy;
% % Korea
the National Science Foundation, the Korea National Research Foundation (No. NRF 2022R1A3B1078756), Korea;
% MExico
CONAHCyT for supporting the national projects CBF2023-2024-427, and CF-2023-G-643., Mexico;
% Morocco
%, Morocco;
% Poland
Ministry of Science and Higher Education (2022/WK/15) and the National Science Centre (UMO-2022/46/E/ST2/00336), Poland;
% Russia
% Spain
MICIU, Spain and NextGenerationEU/PRTR, EU, Spain;
% Sweden
%, Sweden;
% Switzeland
ETHZ, SERI and SNSF, Switzerland;
% UK
STFC and UKRI, UK;
% Ukraine
%, Ukraine;
% US
We also thank CERN for use of the Neutrino Platform.
% Computing
% Individuals
In addition, the participation of individual researchers and institutions has been further supported by funds from  
% Czechia Republic
Charles University (Grant PRIMUS 23/SCI/025 and Research Center UNCE/24/SCI/016), Czech Republic;
%Europe
 H2020-MSCA-RISE-2019 SK2HK no 872549 and H2020 Grant No. RISE-GA822070-JENNIFER2 2020, Europe;
 % Japan
 JSPS KAKENHI Grant JP24K17065; 
 % Korea
National Research Foundation of Korea (NRF-2022R1A5A1030700, NRF-2022R1A3B1078756) funded by the Ministry of Science, Information and Communication Technology (ICT) and the Ministry of Education (RS-2024-00442775), Korea;
% MExico
Associate Dean of Research and Scientific Graduate Studies, Dr. Daniel A. Jacobo-Velázquez, for his support and also to CONAHCyT for the postgraduate scholarship 835328; Rector of the CUCEI, Dr. Marcos Pérez, for his financial and logistical support and CONAHCyT for the information technologies doctoral scholarship 792151, Mexico;
% Poland
Ministry of Science and Higher Education, Republic of Poland, "International co-financed projects", grant number 5316/H2020/2022/2023/2, Poland;
% Spain
LSC funds from MICIU, DGA and UZ, Spanish Ministry of Science and Innovation PID2022-136297NB-I00 /AEI/10.13039/501100011033/ FEDER, UE; CERCA program of the Generalitat de Catalunya; Plan de Doctorados Industriales of the Research and Universities Department of the Catalan Government (2022 DI 011); MICIIN; European Union NextGenerationEU(PRTR-C17.I1); Generalitat de Catalunya;
Spanish Ministry of Innovation and Science under grants MCINN-23-PID2022-139198NB-I00 and PID2021-124050NB-C31, Spain;
% Switzerland
SNF 20FL20-216674, SNF 200021L-231581 and SNF PCEFP2-203261, Switzerland;
% UK
the Leverhulme Trust Research Fellowship Scheme and the 
UKRI Future Leaders Fellowship grant number MR/S034102/1, UK.

\bibliography{reference}

\providecommand{\newblock}{}
\begin{thebibliography}{10}
\expandafter\ifx\csname url\endcsname\relax
  \def\url#1{{\tt #1}}\fi
\expandafter\ifx\csname urlprefix\endcsname\relax\def\urlprefix{URL }\fi
\providecommand{\eprint}[2][]{\url{#2}}

\bibitem{SK}
Fukuda Y {\em et~al.\/} (Super-Kamiokande) 1998 {\em Phys. Rev. Lett.\/} {\bf 81} 1562--1567 (\textit{Preprint} \eprint{hep-ex/9807003})

\bibitem{T2KExperiment}
Abe K {\em et~al.\/} (T2K) 2011 {\em Nucl. Instrum. Meth. A\/} {\bf 659} 106--135 (\textit{Preprint} \eprint{1106.1238})

\bibitem{NA61:2014lfx}
Abgrall N {\em et~al.\/} (NA61) 2014 {\em JINST\/} {\bf 9} P06005 (\textit{Preprint} \eprint{1401.4699})

\bibitem{Nagai:2810696}
Nagai Y (NA61/SHINE) 2022 {Additional Information concerning the Low Energy Beam project} Tech. rep. CERN Geneva \urlprefix\url{https://cds.cern.ch/record/2810696}

\bibitem{Georgi:1974sy}
Georgi H and Glashow S~L 1974 {\em Phys. Rev. Lett.\/} {\bf 32} 438--441

\bibitem{K2K}
Ahn M {\em et~al.\/} (K2K) 2006 {\em Phys. Rev. D\/} {\bf 74} 072003 (\textit{Preprint} \eprint{hep-ex/0606032})

\bibitem{CERN-ESU-015}
 2020 {2020 Update of the European Strategy for Particle Physics (Brochure)} Tech. rep. Geneva \urlprefix\url{https://cds.cern.ch/record/2721370}

\bibitem{Bronner:2020ibs}
Bronner C, Nishimura Y, Xia J and Tashiro T 2020 {\em J. Phys. Conf. Ser.\/} {\bf 1468} 012237

\bibitem{Botao:2867639}
{Letter of Intent: The Hyper-K Underwater Electronics Assembly project (CERN-SPSC-2023-021 ; SPSC-I-260)} \url{https://cds.cern.ch/record/2867639}

\bibitem{Botao:2887744}
{Addendum to the Letter of Intent CERN-SPSC-2023-021 ; SPSC-I-260 (CERN-SPSC-2024-004 ; SPSC-M-796)} \url{https://cds.cern.ch/record/2887744}

\bibitem{BeamUpgrade}
Ishida T 2013 {T2HK: J-PARC upgrade plan for future and beyond T2K} (\textit{Preprint} \eprint{1311.5287})

\bibitem{giganti_2024_12704703}
Giganti C 2024 T2k experiment status and plans \urlprefix\url{https://doi.org/10.5281/zenodo.12704703}

\bibitem{T2K:2019bcf}
Abe K {\em et~al.\/} (T2K) 2020 {\em Nature\/} {\bf 580} 339--344 [Erratum: Nature 583, E16 (2020)] (\textit{Preprint} \eprint{1910.03887})

\bibitem{sfgd-seminar-cern}
Sgalaberna D 2024 {CERN Detector seminar: The Super Fine-Grained Detector (SuperFGD) for the T2K neutrino oscillation experiment} \url{https://indico.cern.ch/event/1484063/}

\bibitem{ND280upgrade-tdr}
Abe K {\em et~al.\/} (T2K) 2019  (\textit{Preprint} \eprint{1901.03750})

\bibitem{Giganti:2713578}
Giganti C~L~P, Lux T~I~B and Yokoyama M~U~o~T (T2K) 2020 {NP07: ND280 Upgrade project} Tech. rep. CERN Geneva \urlprefix\url{https://cds.cern.ch/record/2713578}

\bibitem{Blondel:2017orl}
Blondel A {\em et~al.\/} 2018 {\em JINST\/} {\bf 13} P02006 (\textit{Preprint} \eprint{1707.01785})

\bibitem{Attie:2022smn}
Atti\'e D {\em et~al.\/} 2023 {\em Nucl. Instrum. Meth. A\/} {\bf 1052} 168248 (\textit{Preprint} \eprint{2212.06541})

\bibitem{tpc-seminar-cern}
Levorato S 2024 {Technical challenges for the new T2K High Angle TPCs} \url{https://indico.cern.ch/event/1431318/}

\bibitem{Korzenev:2021mny}
Korzenev A {\em et~al.\/} 2022 {\em JINST\/} {\bf 17} P01016 (\textit{Preprint} \eprint{2109.03078})

\bibitem{bhadra2014letterintentconstructnuprism}
Bhadra S {\em et~al.\/} 2014 Letter of intent to construct a nuprism detector in the j-parc neutrino beamline (\textit{Preprint} \eprint{1412.3086}) \urlprefix\url{https://arxiv.org/abs/1412.3086}

\bibitem{protocollaboration2018hyperkamiokandedesignreport}
Abe K {\em et~al.\/} 2018 Hyper-kamiokande design report (\textit{Preprint} \eprint{1805.04163}) \urlprefix\url{https://arxiv.org/abs/1805.04163}

\bibitem{Barbi:2712416}
Barbi M {\em et~al.\/} 2020 {Proposal for A Water Cherenkov Test Beam Experiment for Hyper-Kamiokande andFuture Large-scale Water-based Detectors} Tech. rep. CERN Geneva \urlprefix\url{https://cds.cern.ch/record/2712416}

\bibitem{T2K:2024wfn}
Abe K {\em et~al.\/} (T2K, Super-Kamiokande) 2025 {\em Phys. Rev. Lett.\/} {\bf 134} 011801 (\textit{Preprint} \eprint{2405.12488})

\bibitem{Super-Kamiokande:2023ahc}
Wester T {\em et~al.\/} (Super-Kamiokande) 2024 {\em Phys. Rev. D\/} {\bf 109} 072014 (\textit{Preprint} \eprint{2311.05105})

\bibitem{Hyper-Kamiokande:2021frf}
Abe K {\em et~al.\/} (Hyper-Kamiokande) 2021 {\em Astrophys. J.\/} {\bf 916} 15 (\textit{Preprint} \eprint{2101.05269})

\bibitem{Super-Kamiokande:2023jbt}
Abe K {\em et~al.\/} (Super-Kamiokande) 2024 {\em Phys. Rev. D\/} {\bf 109} 092001 (\textit{Preprint} \eprint{2312.12907})

\bibitem{Super-Kamiokande:2022ncz}
Abe K {\em et~al.\/} (Super-Kamiokande) 2023 {\em Phys. Rev. Lett.\/} {\bf 130} 031802 [Erratum: Phys.Rev.Lett. 131, 159903 (2023)] (\textit{Preprint} \eprint{2209.14968})

\bibitem{T2K:2017ega}
Abe K {\em et~al.\/} (T2K) 2017 {\em Phys. Rev. D\/} {\bf 95} 111101 (\textit{Preprint} \eprint{1703.01361})

\bibitem{T2K:2019jwa}
Abe K {\em et~al.\/} (T2K) 2019 {\em Phys. Rev. D\/} {\bf 100} 052006 (\textit{Preprint} \eprint{1902.07598})

\bibitem{T2K:2014xvp}
Abe K {\em et~al.\/} (T2K) 2015 {\em Phys. Rev. D\/} {\bf 91} 051102 (\textit{Preprint} \eprint{1410.8811})

\bibitem{Yeh:2011zz}
Yeh M, Hans S, Beriguete W, Rosero R, Hu L, Hahn R~L, Diwan M~V, Jaffe D~E, Kettell S~H and Littenberg L 2011 {\em Nucl. Instrum. Meth. A\/} {\bf 660} 51--56

\bibitem{Franks:2023ttv}
Franks M {\em et~al.\/} 2024 {\em Eur. Phys. J. C\/} {\bf 84} 202 (\textit{Preprint} \eprint{2309.03131})

\bibitem{Buhrer:2024sio}
B\"uhrer N, Alonso-Monsalve S, Franks M, Dieminger T and Sgalaberna D 2024  (\textit{Preprint} \eprint{2410.10519})

\bibitem{Berns:2020ehg}
Berns S, Boyarintsev A, Hugon S, Kose U, Sgalaberna D, De~Roeck A, Lebedynskiy A, Sibilieva T and Zhmurin P 2020 {\em JINST\/} {\bf 15} 10 (\textit{Preprint} \eprint{2011.09859})

\bibitem{3DET:2022dkw}
Berns S {\em et~al.\/} (3DET) 2022 {\em JINST\/} {\bf 17} P10045 (\textit{Preprint} \eprint{2202.10961})

\bibitem{Weber:2023joy}
Weber T {\em et~al.\/} 2023  (\textit{Preprint} \eprint{2312.04672})

\bibitem{Boyarintsev:2021uyw}
Boyarintsev A {\em et~al.\/} 2021 {\em JINST\/} {\bf 16} P12010 (\textit{Preprint} \eprint{2108.11897})

\bibitem{Colaleo:2885937}
Colaleo A {\em et~al.\/} 2024  \urlprefix\url{https://cds.cern.ch/record/2885937}

\bibitem{CERN-DRDC-2024-001}
Sajan E {\em et~al.\/} 2023  \urlprefix\url{https://cds.cern.ch/record/2884872}

\bibitem{3det}
{3D printed detectors (3DET) R\&D collaboration} \urlprefix\url{https://threedet.web.cern.ch}

\bibitem{NA61SHINE:2016nlf}
Abgrall N {\em et~al.\/} (NA61/SHINE) 2016 {\em Eur. Phys. J. C\/} {\bf 76} 617 (\textit{Preprint} \eprint{1603.06774})

\bibitem{NA61SHINE:2018rhe}
Abgrall N {\em et~al.\/} (NA61/SHINE) 2019 {\em Eur. Phys. J. C\/} {\bf 79} 100 (\textit{Preprint} \eprint{1808.04927})

\bibitem{NA61SHINE:2015bad}
Abgrall N {\em et~al.\/} (NA61/SHINE) 2016 {\em Eur. Phys. J. C\/} {\bf 76} 84 (\textit{Preprint} \eprint{1510.02703})

\bibitem{Aduszkiewicz:2309890}
Aduszkiewicz A (NA61/SHINE) 2018 {Study of Hadron-Nucleus and Nucleus-Nucleus Collisions at the CERN SPS: Early Post-LS2 Measurements and Future Plans} Tech. rep. CERN Geneva \urlprefix\url{https://cds.cern.ch/record/2309890}

\end{thebibliography}

\end{document}